\documentclass[apj]{emulateapj}

\def\simlt{\lower.5ex\hbox{$\; \buildrel < \over \sim \;$}}
\def\simgt{\lower.5ex\hbox{$\; \buildrel > \over \sim \;$}}

\def\hnot{\ifmmode H_0 \else H$_0$\fi}

\def\msun{\ifmmode {M_\odot} \else $M_\odot$\fi}
\def\lsun{\ifmmode {L_\odot} \else $L_\odot$\fi}

\def\deg{\ifmmode ^{\circ}
         \else $^{\circ}$\fi}
\def\pdeg{\ifmmode
           $\setbox0=\hbox{$^{\circ}$}\rlap{\hskip.11\wd0 .}$^{\circ}
     \else \setbox0=\hbox{$^{\circ}$}\rlap{\hskip.11\wd0 .}$^{\circ}$\fi}
\def\msunyr{\ifmmode {\rm M_\odot~yr^{-1}}\else${\rm M_\odot~yr^{-1}}$\fi}
\def\lam{\ifmmode {\lambda} \else {$\lambda$} \fi}
\def\lamLlam{\ifmmode \lambda L_{\lambda}(5100) \else {$\lambda L_{\lambda}(5100)$} \fi}
\def\nuLnu{\ifmmode \nu L_{\nu}(5100) \else {$\nu L_{\nu}(5100)$} \fi}

\def\mdoto{\ifmmode {\dot{M}_0} \else  $\dot{M}_0$ \fi}
\def\teff{\ifmmode {T_{eff}} \else $T_{eff}$ \fi}
\def\ilam{\ifmmode {I_\lambda} \else  $I_\lambda$ \fi}
\def\flam{\ifmmode {F_\lambda} \else  $F_\lambda$ \fi}
\def\inu{\ifmmode {I_\nu} \else  $I_\nu$ \fi}
\def\fnu{\ifmmode {F_\nu} \else  $F_\nu$ \fi}
\def\yr{\ifmmode {\rm yr} \else  yr \fi}
\def\cm{\ifmmode {\rm cm} \else  cm \fi}
\def\cmmitwo{\ifmmode \rm cm^{-2} \else $\rm cm^{-2}$\fi}
\def\cmmithree{\ifmmode \rm cm^{-3} \else $\rm cm^{-3}$\fi}
\def\cmps{\ifmmode \rm cm~s^{-1}\else $\rm cm~s^{-1}$\fi}
\def\cmpsps{\ifmmode \rm cm~s^{-2}\else $\rm cm~s^{-2}$\fi}
\def\kmps{\ifmmode \rm km~s^{-1}\else $\rm km~s^{-1}$\fi}
\def\kmpspmpc{\ifmmode \rm km~s^{-1}~Mpc^{-1} \else
    $\rm km~s^{-1}~Mpc^{-1}$\fi}
\def\ergps{\ifmmode \rm erg~s^{-1} \else $\rm erg~s^{-1}$ \fi}
\def\ergpspcm{\ifmmode \rm erg~s^{-1}~cm^{-2} \else $\rm erg~s^{-1}~cm^{-2}$ \fi}
\def\ergpspcmphz{\ifmmode \rm erg~s^{-1}~cm^{-2}~Hz^{-1} \else $\rm
erg~s^{-1}~cm^{-2}~Hz^{-1}$ \fi}
\def\ergpspcmpa{\ifmmode \rm erg~s^{-1}~cm^{-2}~\AA^{-1} \else $\rm
erg~s^{-1}~cm^{-2}~\AA^{-1}$ \fi}
\def\ergpsphz{\ifmmode \rm erg s^{-1} Hz^{-1} \else
   $\rm erg s^{-1} Hz^{-1}$ \fi}
\def\mbh{\ifmmode M_{\mathrm{BH}} \else $M_{\mathrm{BH}}$\fi}
\def\msigma{\ifmmode M_{\sigma} \else $M_{\sigma}$\fi}
\def\dmsigma{\ifmmode \delta_{M\sigma} \else $\delta_{M\sigma}$\fi}
\def\mbulge{\ifmmode M_{\mathrm{bulge}} \else $M_{\mathrm{bulge}}$\fi}
\def\mgal{\ifmmode M_{\mathrm{gal}} \else $M_{\mathrm{gal}}$\fi}
\def\lgal{\ifmmode L_{\mathrm{gal}} \else $L_{\mathrm{gal}}$\fi}
\def\lbulge{\ifmmode L_{\mathrm{bulge}} \else $L_{\mathrm{bulge}}$\fi}
\def\mgalstar{\ifmmode M^*_{\mathrm{gal}} \else $M^*_{\mathrm{gal}}$\fi}
\def\mbhstar{\ifmmode M^*_{\mathrm{BH}} \else $M^*_{\mathrm{BH}}$\fi}

\def\mbhsigstar{\ifmmode M_{\mathrm{BH}} - \sigma_* \else $M_{\mathrm{BH}} - \sigma_*$ \fi}
\def\deltalogmbh{\ifmmode \Delta~{\mathrm{log}}~M_{\mathrm{BH}} \else $\Delta$~log~$M_{\mathrm{BH}}$\fi}

\def\sigstar{\ifmmode \sigma_* \else $\sigma_*$\fi}
\def\sigthree{\ifmmode \sigma_{\mathrm{[O~III]}} \else $\sigma_{\mathrm{[O~III]}}$\fi}
\def\sigtwo{\ifmmode \sigma_{\mathrm{[O~II]}} \else $\sigma_{\mathrm{[O~II]}}$\fi}
\def\signl{\ifmmode \sigma_{\mathrm{NL}} \else $\sigma_{\mathrm{NL}}$\fi}
\def\wthree{\ifmmode {\rm FWHM({[O~III]})} \else $FWHM({[O~III]})$ \fi}
\def\wtwo{\ifmmode {\rm FWHM({[O~II]})} \else $FWHM({[O~II]})$ \fi}
\def\mthree{\ifmmode M_{\mathrm [O~III]} \else $M_{\mathrm [O~III]}$ \fi}
\def\mtwo{\ifmmode M_{\mathrm [O II]} \else $M_{\mathrm [O II]}$ \fi}
\def\hbeta{\ifmmode {\rm H}\beta \else H$\beta$\fi}
\def\hbetan{\ifmmode {\rm H}\beta_{\rm n} \else H$\beta_{\rm n}$\fi}

\def\lbreak{\ifmmode L_{\mathrm{break}} \else $L_{\mathrm{break}}$\fi}
\def\lcut{\ifmmode L_{\mathrm{cut}} \else $L_{\mathrm{cut}}$\fi}
\def\led{\ifmmode L_{\mathrm{Ed}} \else $L_{\mathrm{Ed}}$\fi}
\def\lbol{\ifmmode L_{\mathrm{bol}} \else $L_{\mathrm{bol}}$\fi}

\def\etal{~et al.}

\def\mgii{\ifmmode {\rm Mg{\sc ii}} \else Mg~{\sc ii}\fi}
\newcommand{\oiii}{{\sc [O~iii]}}
\newcommand{\oii}{{\sc [O~ii]}}
\newcommand{\sii}{{\sc [S~ii]}}
\newcommand{\feii}{Fe~{\sc ii}}

\slugcomment{Draft version December 18, 2006}

\shorttitle{The Black Hole - Bulge Relationship for QSOs}
\shortauthors{Salviander et al.}

\begin{document}

\title{The Black Hole Mass - Galaxy Bulge Relationship for QSOs in the SDSS DR3}

\author{S. Salviander, G. A. Shields, and K. Gebhardt}
\affil{Department of Astronomy, University of Texas, Austin, TX 78712}

\and

\author{E.W. Bonning}
\affil{Laboratoire de l'Univers et de ses Th\'{e}ories, Observatoire de Paris, F-92195 Meudon Cedex, France}

\begin{abstract}

We investigate the relationship between black hole mass, \mbh, and host galaxy velocity dispersion, \sigstar, for QSOs in Data Release 3 of the Sloan Digital Sky Survey. We derive \mbh\ from the broad \hbeta\ line width and continuum luminosity, and the bulge stellar velocity dispersion from the \oiii\ narrow line width (\sigthree). At higher redshifts, we use \mgii\ and \oii\ in place of \hbeta\ and \oiii. For redshifts $z < 0.5$, our results agree with the \mbhsigstar\ relationship for nearby galaxies. For $0.5 < z < 1.2$, the \mbhsigstar\ relationship appears to show evolution with redshift in the sense that the bulges are too small for their black holes. However, we find that part of this apparent trend can be attributed to observational biases, including a Malmquist bias involving the QSO luminosity. Accounting for these biases, we find $\sim0.2$~dex evolution in the \mbhsigstar\ relationship between now and redshift $z \approx 1.$

\end{abstract}

\keywords{galaxies: active --- quasars: general --- black hole physics}

\section{Introduction}

In the nucleus of nearly every local galactic bulge lies a supermassive black hole, the relic of previous active galactic nucleus (AGN) activity (see review by Kormendy \& Gebhardt 2001). The mass \mbh\ correlates with bulge luminosity (Magorrian et al. 1998) and stellar velocity dispersion, \sigstar\ (Gebhardt et al. 2000a; Ferrarese \& Merritt 2000). This relationship, established for nearby galaxies with $10^6-10^9$ \msun\ black holes, is given by Tremaine et al. (2002) as
\begin{equation}
\label{e:tremaine}
\mbh = (10^{8.13}~\msun)(\sigstar/200~\kmps)^{4.02}.
\end{equation}
The cause of this tight relationship is not well understood. A popular class of models envisions that black hole growth and star formation proceed simultaneously, obscured by surrounding gas until the black hole becomes massive enough to support a luminosity capable of blowing away the fueling gas and halting star formation (Silk \& Rees 1998, Fabian 1999; King 2003). Di Matteo et al. (2005) perform numerical simulations in which energy deposition from the luminosity of the quasi-stellar object (QSO) leads to ejection of residual gas, simultaneously shutting down black hole growth and star formation. With a simple assumption about the heating efficiency, they achieve a tight \mbhsigstar\ relationship. This relationship can also be reproduced by momentum-driven outflow (Murray et al. 2005). Begelman \& Nath (2005) argue that details of the accretion flow near the hole may be important for the \mbhsigstar\ relationship. 

Study of the \mbhsigstar\ relationship for quasars with large look-back times should give information about the comparative evolution of black holes and their host galaxies. This involves measuring the mass of the black hole and the mass (or velocity dispersion) of the host galaxy as a function of cosmic time. Shields et al. (2003, hereafter ``S03'') use \sigthree\ $\equiv$ FWHM(\oiii)$/2.35$, together with \mbh\ derived from the width of the broad \hbeta\ emission line, to investigate the \mbhsigstar relationship in QSOs to a redshift of $z = 3.3$. They define the ``\oiii\ mass'' to be the black hole mass calculated using Equation~\ref{e:tremaine} with \sigthree\ in place of \sigstar. They find good agreement in the mean with the measured \mbh, which suggests little change in the \mbhsigstar\ relationship since redshifts $z$ = 2 to 3. However, S03 had few objects at high $z$ and none in the redshift range $0.33 < z < 1.1$.

In this paper, we extend the work of S03 using the Sloan Digital Sky Survey Data Release 3 (SDSS DR3; Abazajian et al. 2005). The \oiii\ lines remain in the SDSS spectral window up to $z \sim 0.8$. For higher redshifts, out to redshift of $z \sim1.2$, we use \oii ~$\lambda 3727$ for \sigstar\ and \mgii ~$\lambda 2798$ in place of \hbeta. We examine the use of \oiii\ and \oii\ as surrogates for \sigstar, including the effects of emission-line asymmetry, \feii\ emission, and radio luminosity on the \oiii\ and \oii\ widths.

All values of luminosity used in this study are calculated using the cosmological parameters $\hnot = 70~\kmpspmpc, \Omega_{\rm M} = 0.3$, and $\Omega_{\Lambda} = 0.7$.

\section{Methodology}

\subsection{Deriving Black Hole Masses}\label{s:mbh}

The method for calculating black hole masses is summarized in S03. If the BLR gas orbits the black hole, then $\mbh = v^2R/G$. The appropriate velocity is parametrized as $v = f \times$ FWHM, where FWHM is the full width at half maximum of the broad line used, and $f$ depends on the geometry and kinematics assumed for the BLR (McLure \& Dunlop 2001). Some authors assume $f = \sqrt3/2$ for random orbits (Netzer 1990). The BLR radius, derived from echo mapping studies, increases as a function of the continuum luminosity, $R \propto L^{\gamma}$ with $\gamma = 0.5 - 0.7$ (Wandel et al. 1999; Kaspi et al. 2000; Kaspi et al. 2005). Kaspi et al. (2005) find that $\gamma = 0.67$ for the optical continuum, while McLure \& Jarvis (2002) find $\gamma = 0.61$. Simple assumptions involving photoionization physics suggest $\gamma = 0.5$ (S03). Bentz et al. (2006) find $\gamma = 0.52\pm0.04$ after correcting the sample of Kaspi et al. (2005) for host galaxy contamination. The name ``photoionization mass'' for this method is used regardless of the $\gamma$ assumed. The difference between these slopes is not critical for our study, as discussed below. We adopt S03's formula
\begin{equation}
\label{e:mbh}
\mbh = (10^{7.69}~\msun)v_{3000}^2 L_{44}^{0.5},
\end{equation}
where $v_{3000} \equiv$ FWHM(\hbeta)/3000 \kmps\ and the BLR continuum luminosity at 5100~\AA\ is $L_{44} \equiv \nuLnu /(10^{44}~\ergps)$. For higher redshift objects where the 5100~\AA\ continuum is redshifted out of the spectral window, we measure the continuum luminosity at 4000~\AA\ and scale it to 5100~\AA\ by assuming a power law function fitted by Vanden Berk et al. (2001) for SDSS quasar composite spectra, $\fnu \propto \nu^{\alpha_\nu}$ with $\alpha_\nu = -0.44\pm0.1$. The uncertainty has only a $\sim 1\%$ affect on the derived \mbh\ when \lamLlam\ is inferred from the flux at $\lambda4000$. The adopted mass formula (S03) is based on the $\gamma = 0.5$ fit in Figure~6 of Kaspi et al. (2000) and $f = \sqrt3/2$. Onken et al. (2004) obtain a calibration of the black hole mass formula by forcing agreement with the \mbhsigstar\ relationship for Seyfert galaxies. This calibration corresponds to $f^2 = 1.4$, giving masses a factor 1.8 larger than $f^2 = 0.75$. On the other hand, Figure~2 of Kaspi et al. (2005) gives radii a factor 1.8 smaller at $L_{44} = 1$ than in Kaspi et al. (2000) (adjusted for cosmology), which reduces the mass by this factor. Thus, our adopted formula corresponds closely to using the Kaspi et al. (2005) radii and the Onken et al. value of $f$. The recommended fit of Bentz et al. (2006), together with $f = \sqrt3/2$, gives \mbh\ larger by 0.10 dex than Equation \ref{e:mbh}. This constant offset has no effect on the redshift dependence examined here.

\subsection{Surrogates for Bulge Velocity Dispersion}\label{s:sigma}

The velocity dispersion \sigstar\ for host galaxy bulges of AGNs is often difficult to measure because of the luminosity of the active nucleus, and there is interest in possible surrogates for \sigstar. Nelson \& Whittle (1996) show that \sigstar\ in Seyfert galaxies is closely correlated with the width of \oiii\ $\lambda5007$. Nelson (2000) uses \sigthree\ = FWHM(\oiii)$/2.35$ as a surrogate for \sigstar\ in the \mbhsigstar relationship and finds a ($1\sigma$) dispersion of 0.2 dex. If this dispersion represents a real scatter in the relationship, it indicates secondary influences on the width of \oiii. The width of \oiii\ may be affected by outflow (Wilson \& Heckman 1985; Whittle 1992; Nelson \& Whittle), as evidenced by the frequent presence of blue wings on the emission line. Greene \& Ho (2005) find, for a sample of SDSS narrow line AGN with measured \sigstar, that \oiii\ is affected by a blue wing more that \oii\ and \sii. However, Bonning et al. (2005) examined \oiii\ widths in PG QSOs with measured host galaxy luminosity and find that \oiii\ widths agree well in the mean with the \sigstar\ expected on the basis of the Faber-Jackson relation, with a dispersion of 0.13 dex in \sigthree. \oiii\ emission is often blueshifted with respect to lower-ionization lines, such as \oii, which show no average blueshift (or redshift) (Boroson 2005) and as such may be more suitable surrogates for \sigstar.\\ 

\section{Sample and Measurements}

The QSOs in this study were drawn from the SDSS DR3; spectra for all DR3 objects spectroscopically identified by the survey as QSOs within the relevant redshift range were downloaded from the DR3 Catalog Archive Server\footnote{\url{http://cas.sdss.org/astro/en/}}. The Sloan Digital Sky Survey is designed to map one-quarter of the sky (Stoughton et al. 2002). All data are collected with the 2.5-m telescope at the Apache Point Observatory. The SDSS DR3 covers a spectroscopic area of 1360 sq. deg. and contains spectroscopic data for 45,260 QSOs at $z < 2.3$ and an additional 5,767 ``hi-$z$'' QSOs at $z > 2.3$. Photometric data are collected in five different colors, $u, g, r, i,$ and $z$, where $g$ is roughly equivalent to $V$ in the Johnson magnitude system. The target magnitude limit for QSOs is PSF $i < 20.2$. The spectral resolution, $R = 1850$ for the blue channel ($\lambda\lambda$3800 -- 6000) and 2200 for the red channel ($\lambda\lambda$6000 -- 9200), which corresponds to a resolution of 162 \kmps\ and 136 \kmps\ for blue and red, respectively. The spectra are sky subtracted and corrected for telluric absorption features, though some strong night emission lines can persist at $\lambda$5577, $\lambda$6300, and $\lambda$6363 and in the infrared. We corrected the spectra for galactic extinction following Schlegel et al. (1998) and O'Donnell (1994). 

\subsection{Spectrum Measurements}\label{s:model}

Spectrum measurements involved (1) subtracting the optical and ultraviolet Fe II emission blends using empirical templates, (2) fitting the emission lines of interest and measuring the continuum at suitable wavelengths with an automated procedure, and (3) subtracting the instrumental width in quadrature from FWHMs for the lines. An algorithm originally developed by one of the authors (K.G.) to measure stellar absorption features was modified to process the SDSS spectra. The spectra are rebinned in wavelength scale from logarithmic to linear with 1.41~\AA\ pixel$^{-1}$. The continuum flux is measured at $\lambda$4000 and $\lambda$5100, by taking the mean flux over a range of 30 pixels centered on the desired wavelength. Our choice of wavelengths was made in part to avoid strong emission lines. FWHMs were measured for \hbeta, \oiii, the \oii\ $\lambda\lambda3726,3729$ doublet, and the \mgii\ $\lambda\lambda2796,2803$ doublet using least squares fits of Gauss-Hermite functions. These are more suitable than pure Gaussians to model the often peaky cores and asymmetrical wings of AGN emission lines. The expression for a Gauss-Hermite function is
\begin{eqnarray*}
F(x) & = & Ae^{-x^2/2\sigma^2}[1+h_3f_3(x)+h_4f_4(x)]\\
f_3(x) & = & \frac{1}{\sqrt6}(2\sqrt2x^3-3\sqrt2x)\\
f_4(x) & = & \frac{1}{\sqrt{24}}(4x^4-12x^2+3).
\end{eqnarray*}
Here $h_3$ is a measure of the asymmetrical deviation from a Gaussian profile ($h_3 < 0$ indicates a blue wing), and $h_4$ represents the symmetrical deviation from a Gaussian, or the ``boxiness'' of the profile (e.g. $h_4 > 0$ indicates a more peaky rather than boxy profile). The velocity dispersion $\sigma_{GH}$ in the Gauss-Hermite formalism is roughly equivalent to the second moment for the line profile. For a pure Gaussian profile, $\sigma_{GH}$ is exactly equal to the second moment. When the profile deviates from that of a Gaussian due to a blue wing, the $\sigma_{GH}$ of the fit will be smaller than the second moment by about 5\% for a typical values (see below) of $h_3 = -0.1$ and 10\% for a typical $h_4 = 0.1$ (Pinkney et al. 2003). See Van der Marel \& Franx (1993) for further discussion. For this study, we take $\sigma \equiv$ FWHM/2.35, where FWHM is measured from the fits. For typical values of $h_3$, $\sigma_{GH}$ is larger than FWHM/2.35 by 0.05 dex. We examined the use of $\sigma_{GH}$ in the \mbhsigstar\ relationship for QSOs in this study and found no improvement in the scatter. The reliability of the fitting program was verified by fitting artificial spectra with typical line parameters and noise levels. Objects were selected for use based on error criteria described in \S \ref{s:select}, and all spectra of the selected objects were visually inspected for quality of fit and freedom from artifacts.

The \oii\ doublet is rarely resolved in the SDSS spectra, and we took two approaches to measuring the deblended width. (1) The fitting program was used to fit the doublet with the above fitting parameters. We used a fixed intensity ratio $I(\lambda 3729)/I(\lambda3726) = 1.20$ in order to reduce the number of free parameters and the incidence of failed fits. This is based on our measurements of the average intensity ratio of the more widely spaced \sii\ 6717/6731 doublet in 147 low redshift SDSS QSOs. The doublet ratios of \oii\ and \sii\ are similar (see Figure~5.3 of Osterbrock 1989). The uncertainty introduced by this approximation is unimportant for our purposes, based on fits of \oii\ in real and mock spectra using a realistic range of doublet ratios. The rms scatter of the \sii\ ratio is 0.1. (2) Because of the substantial number of failed fits, even for a fixed doublet ratio, we alternatively measured the width of the \oii\ doublet as a single line, which was more often successful. We then derived the intrinsic line width from a calibration curve based on modeling the doublet using two Gaussians over a wide range of widths, using a one-to-one intensity ratio. (Results differed insignificantly for other reasonable intensity ratios such as 1.2.) This procedure was found reliable in tests with simulated spectra; and results agreed for QSOs where both approaches gave successful fits. For our discussion of the \mbhsigstar\ relationship below, we use the results of approach (2) in order to maximize the number of objects. For discussions of line shape parameters, we necessarily use the results of approach (1).

Our profile fits often underestimated the peakiness of the \oiii\ lines, resulting in fits that cropped off the peaky tops of the lines and produced stubbier profiles. Concern over the degree to which this might influence our results led us to measure the widths for 100 of these cropped lines directly using the `splot' package in IRAF \footnote{IRAF is distributed by the National Optical Astronomy Observatories, which are operated by the Association of Universities for Research in Astronomy, Inc., under cooperative agreement with the National Science Foundation.}. Objects with profile fits that were within $\leq 10\%$ of the actual height of the \oiii\ line showed an insignificant discrepancy between our original measurement and the `splot' measurement. For cropping $> 10\%$ of the actual height of the \oiii\ line, the discrepancy in FWHM was more significant, and we excluded these objects from our data sample. This resulted in the removal of $\sim 10$ \% of the objects with otherwise successful measurements of \oiii. Approximately 40\% of the excluded objects had asymmetrical \oiii\ profiles. (See \S~\ref{s:uncert} for further discussion.) 

\begin{figure}[h]
\includegraphics[width=.45\textwidth]{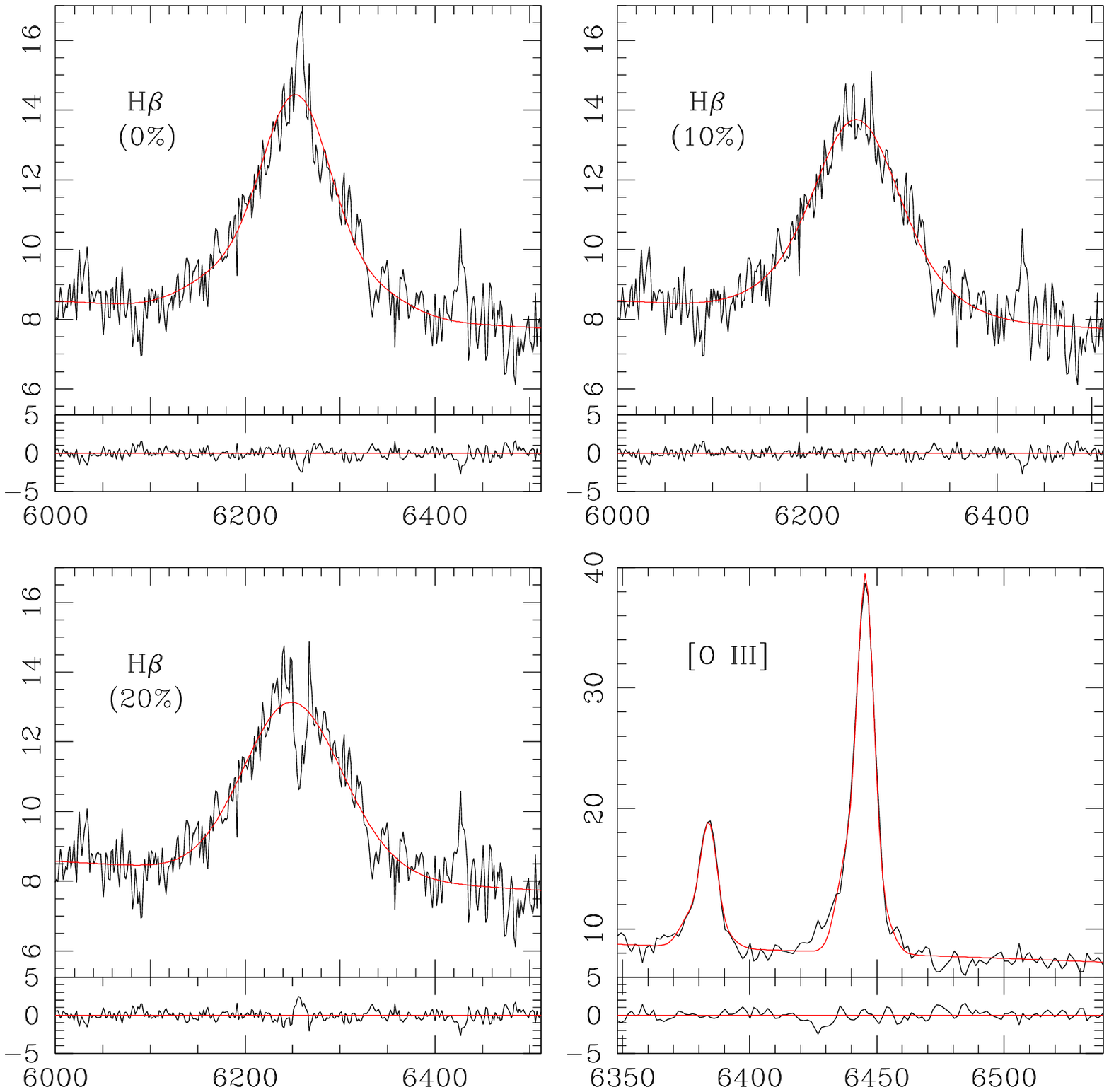}
\includegraphics[width=.45\textwidth]{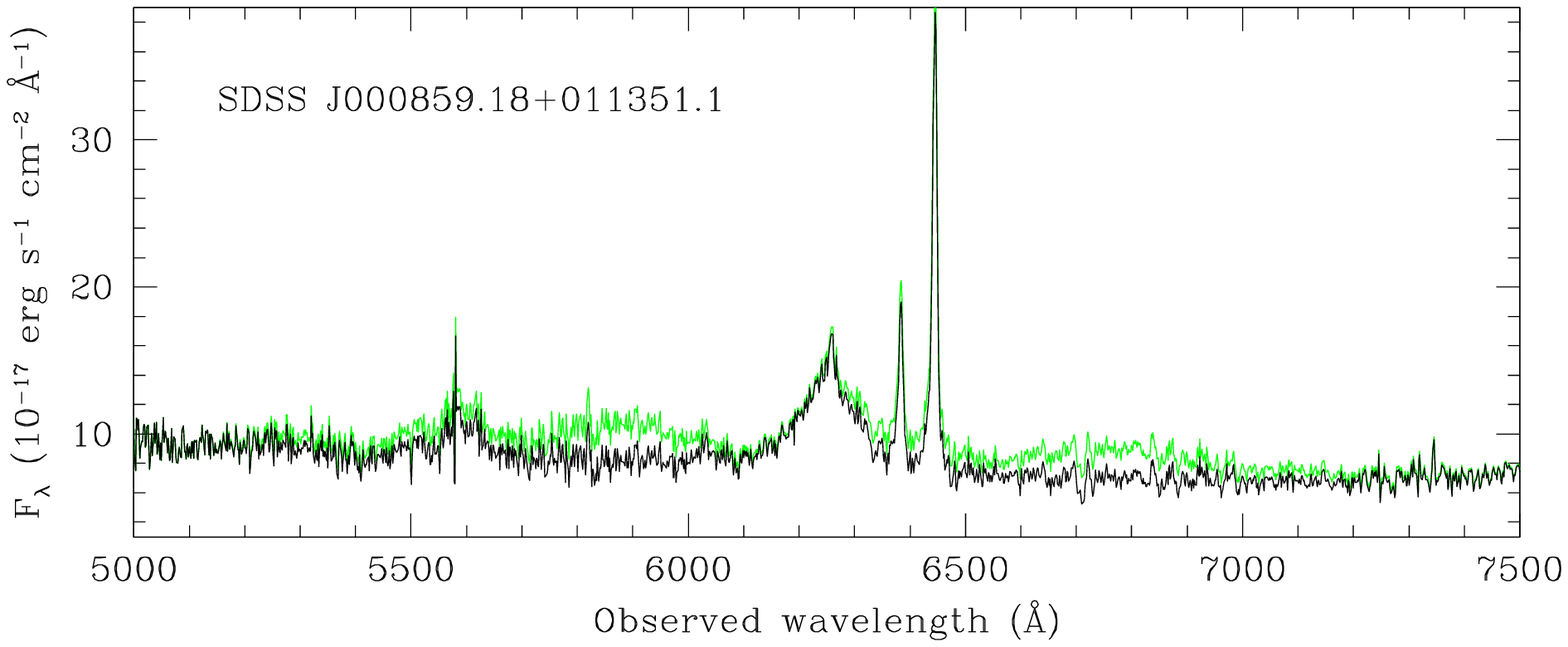}
\figcaption[specfits-ab.eps]{A fit to a representative spectrum of object SDSS J000859.18+011351.1 from the HO3 sample. Lower panel: green and black lines correspond, respectively, to the spectrum prior to and following subtraction of \feii\ emission. Upper panels: red line corresponds to fits to individual emission lines---\oiii, and \hbeta\ (with corrections for the narrow component corresponding to subtraction of 0\%, 10\%, and 20\% of the \oiii\ emission).
\label{f:specfits-ab}}
\end{figure}

\begin{figure}
\includegraphics[width=.45\textwidth]{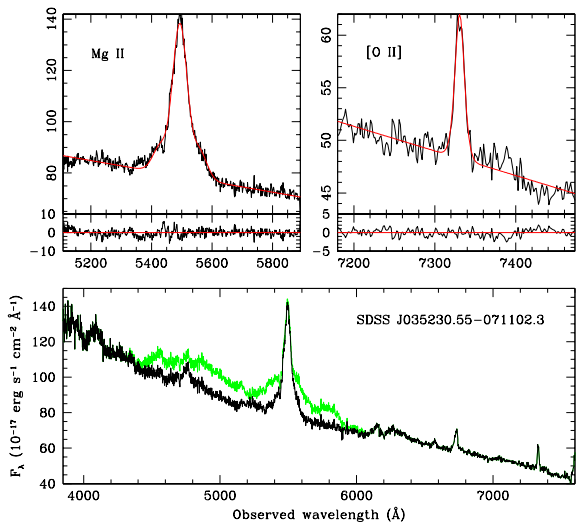}
\figcaption[specfits-c.eps]{A fit to a representative spectrum of object SDSS J035230.55-071102.3 from the MO2 sample. Lower panel: green and black lines correspond, respectively, to the spectrum prior to and following subtraction of \feii\ emission. Upper panels: red lines correspond to fits to individual emission lines---\oii, and \mgii.
\label{f:specfits-c}}
\end{figure}

The profile of \hbeta\ was corrected for the narrow component by subtracting from it an assumed narrow \hbeta\ line (\hbetan)with the profile and redshift of the $\lambda5007$ line but 10\% of its flux. (The \oiii\ $\lambda5007$/\hbetan\ intensity ratio typically is $\sim 10$ for QSOs; Baldwin, Phillips, and Terlevich 1981.) Gauss-Hermite functions model \hbeta\ adequately once this correction has been made, and in most cases a 10\% subtraction was appropriate. We also implemented a 20\% subtraction, or no subtraction at all, for visual comparison. Objects for which the 10\% correction resulted in too little or too much narrow-component subtraction were removed from the sample if the fit to the broad component was affected. This was done for expedience in order to avoid adjusting the fit manually. Very few objects were rejected for over-subtraction of the narrow component, and approximately 14\% of all the removed objects were rejected for having \oiii$/\hbeta_n\ \ll 10$ if this resulted in bad fit to the broad component. It is possible that removing objects on this basis introduces a bias to the sample, in that these objects tend to have low Eddington ratios. See \S \ref{s:uncert} for further discussion. Representative spectrum fits are shown in Figure~\ref{f:specfits-ab} and Figure~\ref{f:specfits-c}.

\subsection{\feii\ Subtraction}\label{s:feii}

Significant difficulties in measuring properties of AGN spectra can arise from blended \feii\ emission lines in both the UV and optical (Boroson \& Green 1992; Vestergaard \& Wilkes 2001). We modeled \feii\ emission in the optical using a template from Marziani et al. (2003), generated from the spectrum of I Zw 1, an object with strong iron emission and relatively narrow ``broad'' emission lines. The template shows two broad blended features in the regions of $\lambda\lambda4450-4700$ and $\lambda\lambda5150-5350$ as well as three strong features in the region of \oiii\ $\lambda\lambda4959,5007$ lines. Vestergaard \& Wilkes present an empirical template for the region of $\lambda\lambda$1250 -- 3090 using the ultraviolet spectrum of I Zw 1. We have reconstructed this template from their Figure~3 in the region of $\lambda\lambda$2180 -- 3060. In the region $\lambda\lambda$2780 -- 2830 centered on \mgii, the Vestergaard \& Wilkes template is set to zero. For this region, we incorporate a theoretical \feii\ model of Sigut \& Pradhan (2003) scaled to match the Vestergaard \& Wilkes template at neighboring wavelengths. In a very few cases the \feii\ in the wings of \mgii\ was too strong for reliable subtraction, and these objects were excluded from our data sample.
 
Modeling of \feii\ emission for each of our spectra is carried out by first convolving the optical template with the FWHM of \hbeta\ and the UV template with the FWHM of \mgii, based on preliminary spectrum fits without \feii\ subtraction. Subtraction of the optical \feii\ was done by scaling and subtracting the template so as to give minimum deviation from a straight line between two 70~\AA\ continuum bands straddling the $\lambda\lambda5150-5350$ \feii\ band. (If necessitated by the redshift, we instead used the $\lambda\lambda4450-4700$ \feii\ band.) Many of our QSOs had redshifts such that the short wavelength end of the ultraviolet \feii\ blend, which extends down to $\lambda2400$, was inaccessible. This made it difficult to fit the entire template to the spectrum or to determine the continuum slope precisely in this spectral region. Therefore, the \feii\ template was scaled and subtracted so as to bring the flux in the region $\lambda\lambda$2980 -- 3020 into agreement with the observed flux in the region $\lambda\lambda$3020 -- 3100, which is relatively devoid of \feii\ emission. We assumed an underlying power-law continuum $F_{\lambda} \propto \lambda^{-1.21}$ (Vanden Berk et al. 2001) to scale the flux to 3000~\AA\ from 3060~\AA. The model was subtracted from the spectra, and the continuum flux and emission line parameters were measured again. All fits were visually inspected following \feii\ subtraction and objects with failed subtractions were eliminated. 

We examined the sensitivity of the UV \feii\ subtraction to various details of our procedure. (1) Increasing the width of the Gaussian used to broaden the template by a factor 2 resulted in a typical change in FWHM(\mgii) of $\pm1$ or 2\%, and a mean decrease of 1\%. This indicates that uncertainties in the \feii\ broad line width introduce insignificant errors. (2) Assuming a slope of $\flam \sim \lambda^{-1.56}$ (i.e., $\fnu \sim \nu^{-0.44}$) instead of $\lambda^{-1.21}$ to predict the flux at 2970 from flux at 3060 increases the flux at 2970 by 1.1\%. This gave a mean decrease in the fitted \mgii\ FWHM by 0.4\% and an rms change of 0.8\%. Thus, uncertainties in the continuum slope in the near ultraviolet, including typical object-to-object variations, have little effect on the post-subtraction width of \mgii. (3) We repeated our fits with the \feii\ template set to zero in the region $\lambda\lambda$2780 -- 2830, rather than interpolating theoretical values as described above. This resulted in a mean increase in \mgii\ FWHM of 1.3\%, and an rms change of 2.2\%.

Although our procedure was insensitive to the details just described, the \feii\ subtraction in about 20\% of the objects appeared to be imperfect, most often under-subtracted. This could result from noise or artifacts or a mismatch of the template to the \feii\ spectrum of the object. In order to assess the degree to which iron subtraction could affect the width of \mgii, we performed the following tests: (1) For a subsample of objects with various amounts of \feii, we scaled up or down the amount of \feii\ measured as much as the data seemed to tolerate in a visual inspection. This was done in particular for a number of QSOs with sufficient redshift to make the entire \feii\ blend down to $\lambda2000$ visible in the spectrum. For most objects, an increase or decrease by 50\% in the iron subtraction was clearly wrong. However, at the level of 20\% increase or decrease of the amount of \feii\ found in the automated procedure, it became difficult to discern an under/over-subtraction from an ``ideal'' subtraction. This factor of 20\%, which we adopt as the $1 \sigma$ error for our iron subtraction, resulted in an average uncertainty of $< 2\%$ in \mgii\ width.
(2) We also compared \feii-subtracted and -unsubtracted \mgii\ FWHMs to determine the effect of neglecting removal of \feii\ altogether, and found that the effect of \feii-subtraction depends on \mgii\ width. For objects with \mgii\ FWHM $< 4000$ \kmps, \feii-subtraction has the effect of increasing \mgii\ FWHM by an average of 7\%, and by as much as 20\%. For objects with \mgii\ FWHM $> 4000$ \kmps\ \feii-subtraction has the effect of decreasing \mgii\ FWHM by similar factors. This result can be understood as a function of how the \mgii\ line profile changes with \feii\ subtraction. 

\subsection{Subsample Selection}\label{s:select}

A lower-redshift subsample was created for study of the \mbhsigstar relationship using \hbeta\ and \oiii\ (hereinafter the ``HO3'' sample). These objects were selected from among the survey QSOs on the basis of redshift alone in order to include the widest possible range of luminosities. This corresponded to a redshift range of $z < 0.81$, which ensures that the \oiii\ emission lines are at least 100~\AA\ inside the SDSS wavelength limits. There are 12,263 DR3 QSOs with $z < 0.81$. A series of quality cuts were made, based on our measurements. Elimination of objects for which \hbeta\ or \oiii\ lines were not fit by our program, due to low S/N or absence of the line, left 11,057 objects. We eliminated objects with (1) \hbeta\ FWHM $< 1500$ \kmps, (2) \oiii\ FWHM $< \sqrt2$ $\times$ instrumental width ($\sim10$\% of all objects with \oiii\ measurements), and EW errors greater than 5\% for (3) \hbeta\ and (4) \oiii. These four numerical cuts reduced the sample to (1) 9869, (2) 9474, (3) 5294, and (4) 5036 objects, respectively. Further cuts were made to eliminate objects with FWHM errors $> 10\%$, absolute value of $h_4 \geq 0.2$, and reduced $\chi^2 > 4$. The $h_4$ cut was implemented to obtain the best fit for the largest number of objects---for the \oii\ emission line in particular, our fitting algorithm would sometimes attempt to fit troughs on either side of the emission line, which exist in the original spectra and are presumably the result of low S/N. This produced an unreasonably large negative value for the $h_4$ parameter and a poor fit to the actual profile. The FWHM error, $h_4$, and $\chi^2$ cuts left 3999, 3749 and 3665 objects, respectively. A visual inspection of the remaining spectra was conducted to remove objects with failed fits, poor S/N, artifacts (e.g. cosmic ray spikes), and grossly irregular line profiles. Less than 10\% of the discarded objects were rejected for irregular line profiles. These consisted mostly of objects with double-peaked \hbeta\ profiles, which may be indicative of SMBHs in low-luminosity, low-accretion states (Strateva et al. 2003; Eracleous \& Halpern 2003). Visual inspections yielded a final sample of 1736 objects.

We have not attempted to subtract the host galaxy contribution to the observed spectrum or to remove objects with strong galaxy contributions (see \S~\ref{s:uncert}).

A second data subsample was created to study the \mbhsigstar\ relationship at higher redshifts using \mgii\ and \oii\ (the ``MO2'' sample). There are 19,011 DR3 QSOs in the redshift range, $0.4 \leq z \leq 1.4$, in which both \mgii\ and \oii\ are accessible. Our program produced fits for 5419 objects. The analogous quality cuts as for the HO3 sample were then applied here with the exception of the EW error cut. The 5\% EW error cut was applied to \mgii, but for \oii\ it yielded very few objects. Therefore we relaxed the EW error cut to 10\% for \oii. These cuts left (1) 5419, (2) 5084, (3) 5084, and (4) 1034 objects (same numbering as for HO3). Further cuts were made on the basis of FWHM errors $> 10\%$, $h_4 \geq 0.2$, and reduced $\chi^2 > 4$, as with the HO3 sample. The FWHM error cut resulted in 384 objects; the $h_4$ cut resulted in 349 objects; the $\chi^2$ cut had no effect; and visual inspections yielded a final sample of 158 objects, with a maximum redshift of $z = 1.19$. Most visual rejections were for noisy spectra that were accidentally well fit by the routine. Only a handful of objects were discarded for irregularities in the line profile shape (not due to low S/N). These consisted of double-peaked \mgii\ lines. Less than 5\% of the objects were discarded for having excessive \feii\ emission. The small number of useful objects largely reflects the weakness of \oii\ in most QSOs.

With so few objects, we should ask whether there is something special about QSOs with measurable \oii\ that could bias our result. One concern is contamination of \oii\ by emission from star-forming regions. High \oii/\oiii\ line ratios are believed to be indicative of this (Ho 2005; Kim et al. 2006), but for our objects with \oii\ and \oiii\ in common, we find that this ratio is not abnormally high. The average \oii/\oiii\ line ratio is 0.12, consistent with pure-AGN emission in Type 1 objects (Kim et al.), with approximately 5\% of our objects showing modestly elevated ratios. Another concern is possible introduction of bias through correlations between occurence of measurable \oii\ and spectral properties. In \S~\ref{s:mbhsigstar} we discuss correlations involving \feii\ emission and Eddington ratio.

\subsection{Radio Loudness}\label{s:radio}

The radio-to-optical flux ratio, or radio loudness, was calculated for each object in the two samples in terms of the parameter log $R = $ log $(F_r/F_o)$, where $F_r$ and $F_o$ are the monochromatic flux \fnu\ in the radio at 5.0 GHz and in the optical at $\lambda4400$ ($B$-band), respectively. Kellerman et al. (1989) define radio-quiet as log $R < 0$, and radio-loud as log $R > 1$; but here we take radio quiet as log $R < 1$. Radio fluxes were obtained from the SDSS Quasar Catalog III (Schneider et al. 2005), which quotes the 20-cm peak flux density (AB magnitude) from the FIRST catalog (White et al. 1997) for most DR3 quasars detected by FIRST as radio sources. For objects not listed in the quasar catalog, radio fluxes were obtained from the FIRST catalog directly. Radio fluxes for objects residing outside of the FIRST coverage area were taken from the NVSS catalog (Condon et al. 1998). All radio fluxes were scaled from 1.4 GHz to 5.0 GHz using a spectral slope of $\alpha = -0.5$, where $F_{\nu} \propto \nu^{\alpha}$. The quasar catalog includes a search radius of 30.0 arcseconds from the central optical source, but we only consider sources within 2.0 arcseconds of the optical source. The flux at $\lambda4400$ was scaled from our measurements of the spectrum at rest wavelength $\lambda5100$ (HO3) or $\lambda3000$ (MO2). Because of the great range in log~$R$ among QSOs, the exact value of continuum slope used for scaling has little affect on the number of objects that qualify as RLQ or on our conclusions.

We found 90 radio-loud QSOs (RLQ) in the HO3 sample and 35 RLQ in the MO2 sample, comprising $\sim~5\%$ and $\sim~22\%$ of the respective samples. Our RLQ ranged up to log $R = 4$ and 5 for the HO3 and MO2 samples, respectively. About one-quarter and one-fifth of the HO3 and MO2 RLQ, respectively, are identified by the catalog as having extended FIRST emission. The percentage of RLQ increases with redshift in the HO3 sample. Compared to the RQQ, the RLQ have larger \mbh\ and more luminous \oiii\ emission lines with less prominent blue wings. There is no consistent tendency for RLQ in the MO2 sample to be associated with greater \mbh\ or higher \oii\ luminosity. For the redshift range in common, the percentage of RLQ in MO2 is twice than in HO3.

\section{Results}

\subsection{The Broad Lines}\label{s:bl}

\begin{figure}[h]
\includegraphics[width=.45\textwidth]{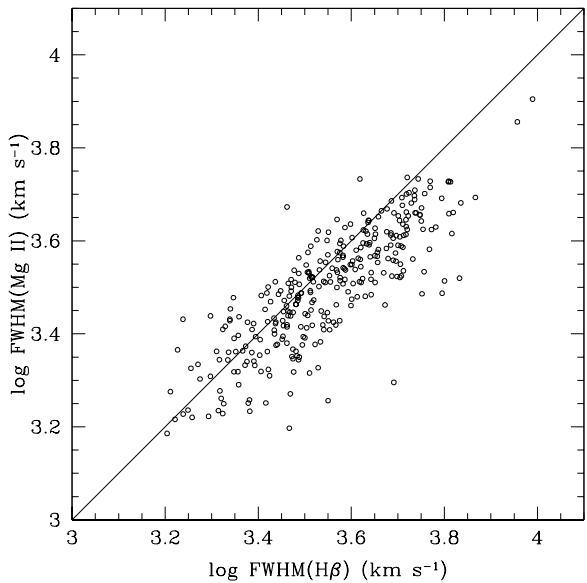}
\figcaption[hbmg-new.eps]{FWHM(\mgii) versus FWHM(\hbeta). The solid line represents a one-to-one relationship, and is not a fit to the data. The mean difference in width is $<$log FWHM(\hbeta) - log FWHM(\mgii)$>$ = 0.05.
\label{f:hbmg}}
\end{figure}

Figure \ref{f:hbmg} shows a strong correlation between \hbeta\ and \mgii\ widths, consistent with McLure \& Jarvis (2002). However, the \hbeta\ widths tend to be larger than the \mgii\ widths for FWHM $> 4000$~\kmps. Using simulated spectra as described above, we conducted a test in which the MO2 objects were given, object by object, \mgii\ profiles identical to \hbeta\ in $\sigma$, $h3$, and $h4$, but with the measured \mgii\ EW and ultraviolet \feii\ strength. Our fitting procedure recovered the input parameters, showing that the deviation of \mgii\ width from \hbeta\ width is not an artifact of the fitting. In some cases, the difference in widths involves an extensive red wing on the \hbeta\ line. We have used FWHM of \mgii\ and \hbeta\ interchangeably in this work, but we discuss below the effect of using a recalibration of \mgii\ width based on Figure \ref{f:hbmg}.

\subsection{The Narrow Lines}\label{s:nl}

\begin{figure}
\includegraphics[width=.45\textwidth]{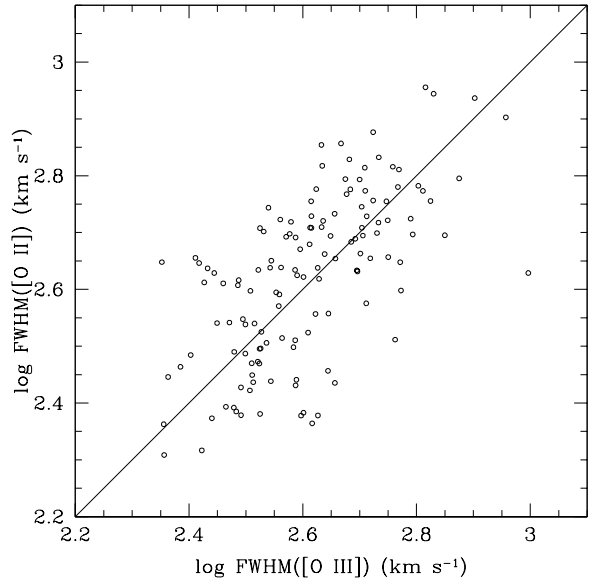}
\figcaption[o3o2-new.eps]{FWHM(\oii) versus FWHM(\oiii). The solid line represents a one-to-one relationship, and is not a fit to the data. The mean difference in width is $<$log FWHM(\oiii) - log FWHM(\oii)$>$ = -0.013.
\label{f:o2o3}}
\end{figure}

Figure \ref{f:o2o3} compares FWHM of \oii\ (determined from single-line modeling described above) and \oiii\ for all objects in the QSO catalog showing both \oii\ and \oiii. In the mean there is good agreement, with log \sigthree\ - log \sigtwo\ = -0.013 and a dispersion of 0.12 dex. (For \oii\ modeled as a doublet, log \sigthree\ - log \sigtwo\ = 0.00 in the mean for the set of objects with successful fits.) The presence of a blue wing, marked by a negative $h_3$, is common for \oiii. The \oii\ lines also show negative $h_3$, but neither as often nor as strongly as \oiii. We find a weak correlation between the Eddington ratio, $\lbol/\led$, and $h_3$ for \oiii\ (but not \oii), in the sense that greater $\lbol/\led$ corresponds to a larger blue wing on \oiii. We take $\led = (1.26 \times 10^{38}~\ergps)(\mbh/\msun$), and $\lbol = 9\nuLnu$ following Kaspi et al. (2005). Greene \& Ho (2005) examine narrow-line AGN in the SDSS DR2 with measured \sigstar. They find that the width of \oii\ is a better tracer of \sigstar\ than \oiii, which on average tends to be wider than \sigstar\ by $\sim0.05$~dex (without the \oiii\ blue wing removed), rather more than the \oii\ vs. \oiii\ offset found here. Our sample consists of Type 1 (broad-line) objects, for which Nelson \& Whittle (1996) find better agreement between \oiii\ width and \sigstar\ than for Type 2 (narrow-line) objects. Some discrepancy may also arise due to differences in modeling of the \oiii\ profiles. For typical values of $h_3$ and $h_4$, $\sigma$ determined by Gauss-Hermite modeling of \oiii\ profiles can be smaller than $\sigma$ measured from the second moment by up to 10\%. Greene \& Ho show that removal of the blue wing of \oiii\ results in a better tracer of \sigstar. We find that the width of \oiii\ (but not \oii) does correlate with $h_3$, but this does not appear to significantly widen the \oiii\ line compared to \oii. We examined the red and blue half widths at half maximum intensity (HWHM) of \oiii\ for our sample. In the mean, the width of the blue side of \oiii\ is larger than the red by 0.026~dex, confirming that a blue wing tends to widen \oiii\ by a modest amount. If the red HWHM is essentially unaffected by the blue component, this suggests that the FWHM on average is increased by 0.013~dex by the blue component. When we compare $2~\times$ HWHM(red) of \oiii\ to the FWHM of \oii\ we find that the average difference is \sigthree\ - \sigtwo\ = -0.028~dex. Consistent with Nelson et al. (2004), we find that the degree of blue asymmetry on \oiii\ is moderately correlated with greater optical \feii\ emission strength, and that the width of \oiii\ is weakly correlated with greater \feii\ strength.

Boroson (2005) finds that a significant blueshift of the \oiii\ line, with respect low-ionization lines such as \oii\ and \sii, correlates with broader \oiii\ line width, though not with blue asymmetry in \oiii. Our results show a moderate correlation between the width of \oiii\ and the \oiii\ blueshift (relative to \oii), especially for objects with blueshift $> 40$ \kmps. Blueshifted \oiii\ correlates weakly with blue asymmetry in \oiii. The average and maximum \oiii\ blueshifts for our sample, relative to \oii, are 15~\kmps\ and 420~\kmps, respectively, compared with 40~\kmps\ and 300~\kmps\ in Boroson (2005).

The kurtosis of a line profile corresponds to the parameter $h_4$, with positive $h_4$ signifying a profile more sharply peaked than a Gaussian. Ninety-five percent of \oiii\ lines show positive $h_4$ , while 80\% of \oii\ lines show positive $h_4$. We find a weak anti-correlation between \oiii\ width and $h_4$, which means that the boxier lines tend to be broader. Conversely, we find a moderate correlation between \oii\ width and $h_4$ such that peakier lines tend to be broader. We find no significant correlation between $h_4$ for \oiii\ or \oii\ and \feii\ strength. There is a weak correlation between $h_4$ and $\lbol/\led$ for \oiii, in the sense that greater $\lbol/\led$ corresponds to peakier \oiii\ profiles. There is no correlation between $h_4$ for \oii\ and $\lbol/\led$. The peakiness of the \oiii\ profiles may be related to dust within the emitting gas clouds (Wilson \& Heckman 1985; Busko \& Steiner 1989); the combination of peakiness and blue asymmetry suggests a dusty NLR with outflow (Netzer \& Laor 1993).

\begin{figure}[h]
\includegraphics[width=.45\textwidth]{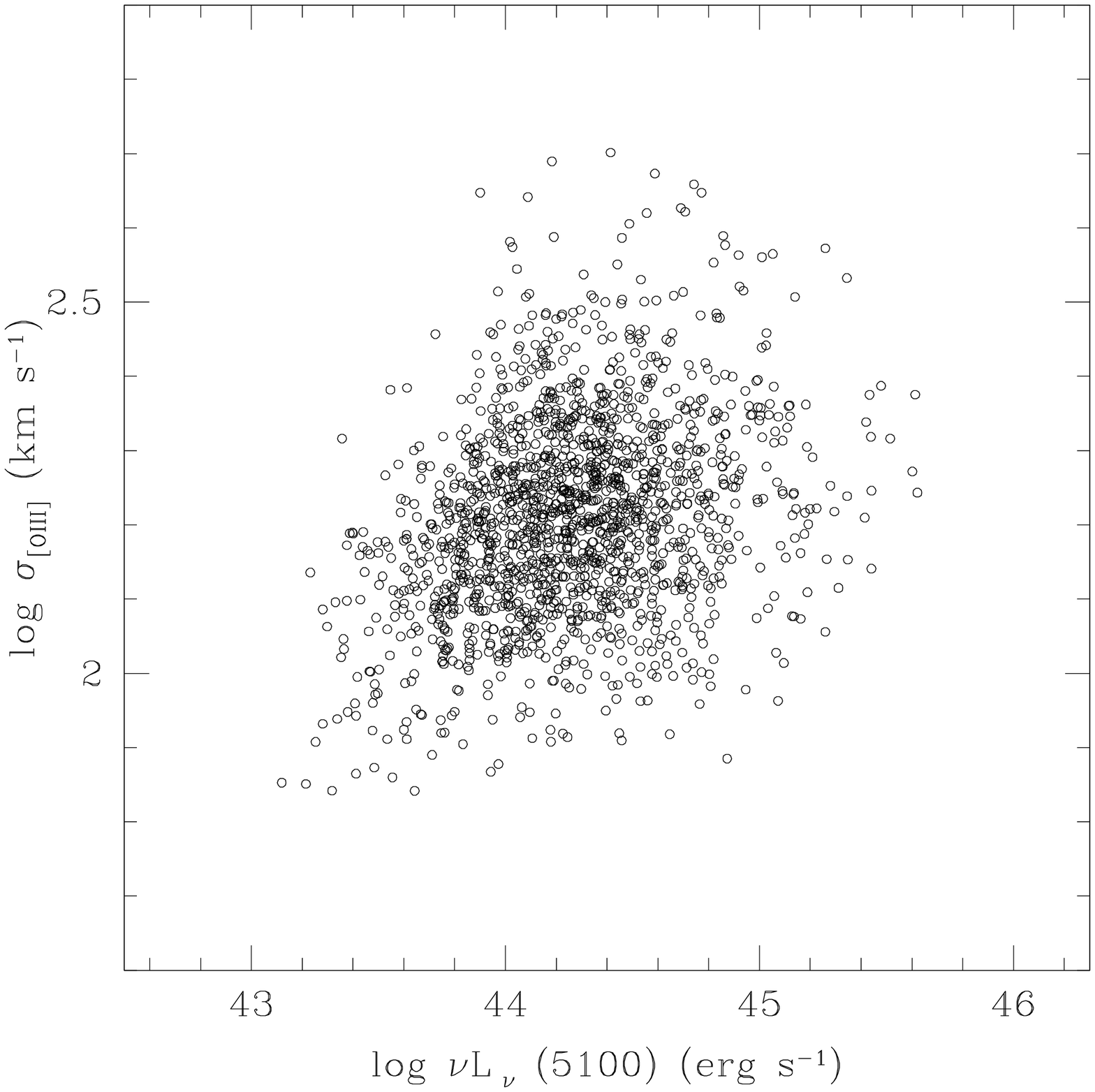}
\figcaption[lum-sigo3-new.eps]{Log \sigthree\ as a function of log \nuLnu(5100~\AA) (excluding radio-loud QSOs).
\label{f:Lo3}}
\end{figure}

\subsection{The Narrow Line Width-Luminosity Relationship}\label{s:vl}

S03 showed that the width of \oiii\ is correlated with \mbh, and Corbett et al. (2003) find a strong relationship between \oiii\ width and QSO luminosity using composite spectra from the 2dF QSO Redshift Survey. However, this relationship is not found by Corbett et al. for other NLR lines, including \oii. We find a weak relationship between continuum luminosity and \oiii\ width for RQQ (Spearman's rank order coefficient $r_S = 0.280$, with probability $P_S < 0.1\%$ that a real correlation is not present), and none between continuum luminosity and \oii\ width. Figure \ref{f:Lo3} shows the width-luminosity relationship for \oiii. Corbett et al. argue that the correlation between \oiii\ width and luminosity suggests that the kinematics of the NLR are affected to some degree by the gravitational potential of the central black hole. However, the correlation could also involve the tendency of bigger black holes (and bigger host galaxies) to be associated with more luminous QSOs. It is curious, then, that a correlation is not found for \oii, given the overall agreement between \sigthree\ and \sigtwo. It is possible that the lack of correlation is due to the limited dynamic range of \oii\ width and luminosity in our sample. When we limited the \oiii\ sample to a comparably-sized width and luminosity range, we found that the correlation between \oiii\ and luminosity disappeared.

\subsection{The \mbhsigstar Relationship}\label{s:mbhsigstar}

\begin{figure}[h]
\includegraphics[width=.45\textwidth]{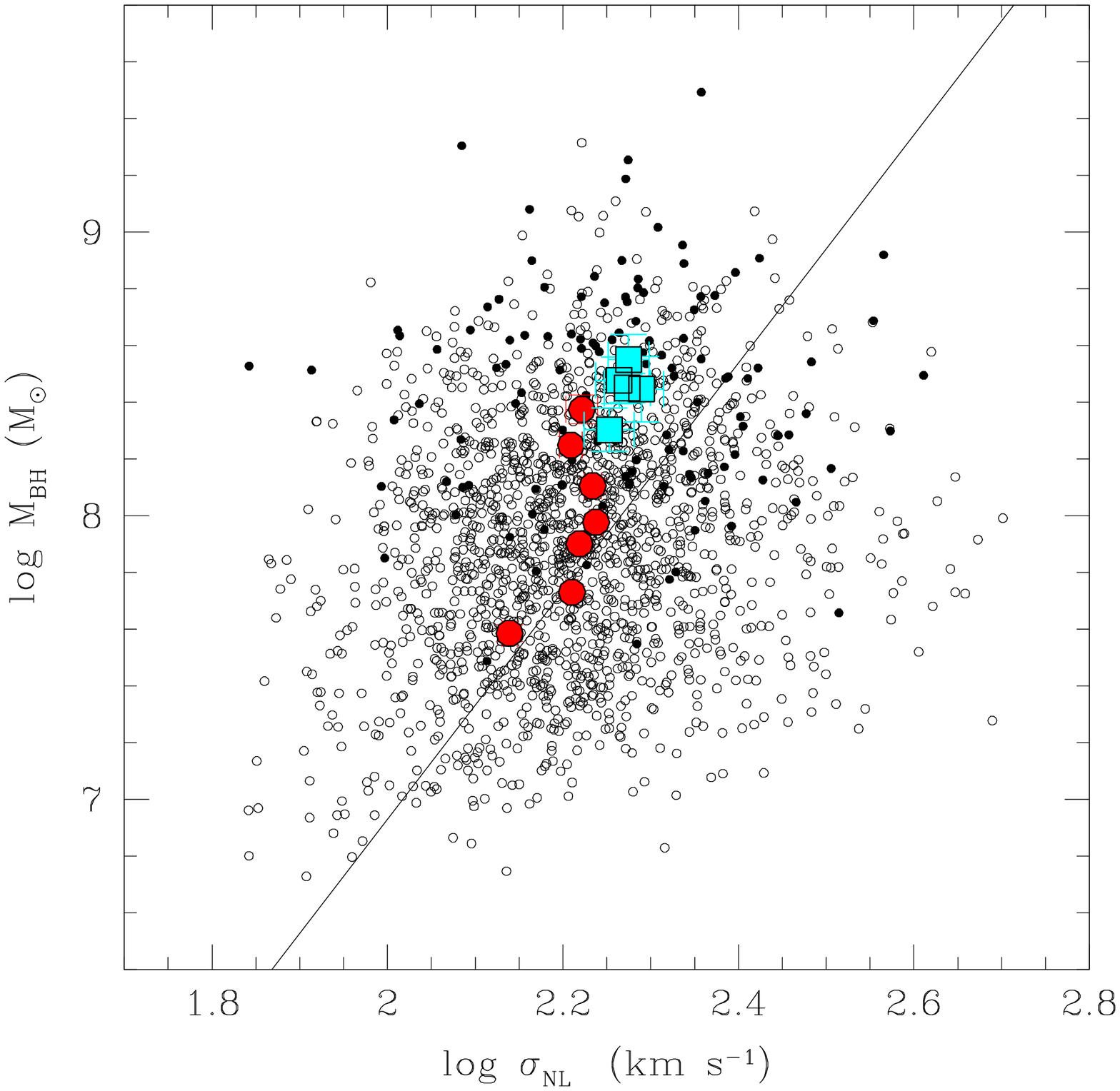}
\figcaption[ ]{The \mbhsigstar\ relationship for our combined sample (excluding radio-loud QSOs). Small open and closed circles denote HO3 and MO2, respectively. For the HO3 sample, \mbh\ is derived from the FWHM of \hbeta\ and the continuum luminosity at 5100~\AA; the velocity dispersion, \signl, is inferred from the FWHM of \oiii. For the MO2 sample, FWHM of \mgii\ and \oii\ are used in place of FWHM of \hbeta\ and \oiii. The continuum luminosity at 4000~\AA\ is scaled to 5100~\AA\ by assuming a power law function, F$_{\nu} \propto \nu^{-0.44}$. The solid line represents the \mbhsigstar\ correlation for nearby galaxies (Equation~\ref{e:tremaine}; Tremaine et al. 2002) and is not a fit to the data. Large red and cyan circles represent mean \mbh\ and \signl\ for HO3 and MO2 in redshift bins $\Delta~z = 0.1$. Error bars show the standard error of the mean. HO3 sample error bars are smaller than the data points. Note that the bins increase monotonically in \mbh\ with redshift for HO3 (see Table~\ref{t:bins}).
\label{f:mbh-both}}
\end{figure}

Figure \ref{f:mbh-both} shows results for the \mbhsigstar\ relationship using both the \oiii\ and \oii\ emission-line widths (the HO3 and MO2 samples, respectively) as surrogates for \sigstar, excluding RLQ. The solid line is not a fit to the data, but rather represents the fit given by Equation~\ref{e:tremaine} (the ``\mbhsigstar\ correlation''). Our results for $z < 0.5$ (see \S5) are consistent with previous findings (Nelson 2000; S03; Boroson 2003; Grupe \& Mathur 2004), with the data points tending to scatter evenly about the \mbhsigstar\ correlation. Because of the large scatter, mostly attributable to the scatter in \sigthree\ as a surrogate for \sigstar, there would be little meaning in fitting a line to the data (see discussion in Boroson 2003). More meaningful is the dispersion with respect to the local \mbhsigstar\ relationship shown in Figure~\ref{f:mbh-both}. This is 0.61 for the HO3 sample (comparable with the findings of S03) and 0.67 for the MO2 sample. Bonning et al. (2005) find a dispersion of 0.13~dex using \sigthree\ as a surrogate for \sigstar\ inferred from host luminosity. Given the $\sigstar^4$ dependence in Equation~\ref{e:tremaine}, the scatter in the \sigthree\ -- \sigstar\ relationship, together with the smaller scatter of 0.3~dex in the \mbhsigstar\ relationship for galaxies (Tremaine et al. 2002), accounts for the scatter in \deltalogmbh\ of 0.61~dex that we find here.

We follow S03 in comparing \mbh\ (from Equation~\ref{e:mbh}) to \msigma, defined as the ``\oiii\ mass'' or ``\oii\ mass'' calculated with Equation~\ref{e:tremaine} using \sigthree\ or \sigtwo\ in place of \sigstar. The mean $\deltalogmbh\ \equiv \mathrm{log}\,\mbh - \mathrm{ log}\,\msigma$ in the HO3 sample is +0.30 and +0.13 for RLQ and RQQ, respectively. The results of Bonning et al. (2005) indicate that this RL -- RQ offset could result from narrower \oiii\ widths for RLQ that underestimate \sigstar\ (see also S03). We find the opposite difference between \deltalogmbh\ for RLQ and RQQ in the MO2 sample, such that the mean \deltalogmbh\ +0.16 for RLQ and +0.45 for RQQ. This may be an indication that \oii\ is not affected by radio loudness in the same way as \oiii. In fact, the difference in widths of \oiii\ and \oii\ is greater for RLQ than for RQQ, with \oii\ tending to be broader than \oiii\ by 0.04 dex for RLQ.

\section{Redshift Dependence}\label{s:zdep}

\begin{figure}[h]
\includegraphics[width=.45\textwidth]{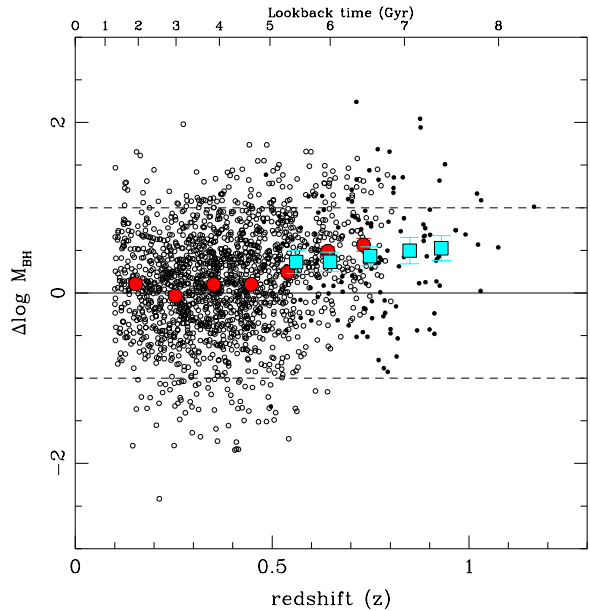}
\figcaption[ ]{Redshift dependence of \deltalogmbh, where $\deltalogmbh = \rm{log}~\mbh - \rm{log}~\msigma$ (excluding radio-loud QSOs). Small open and closed circles denote the HO3 and MO2 samples, respectively. Large red and cyan circles show the mean \deltalogmbh\ in redshift bins $\Delta~z = 0.1$ for the HO3 and MO2 samples, respectively. Error bars show the standard error of the mean. HO3 sample error bars are smaller than the data points.
\label{f:del-both}}
\end{figure}

Does the \mbhsigstar\ relationship evolve with lookback time? We test this by seeing how closely our black hole masses agree with the local \mbhsigstar\ correlation as a function of redshift (see SO3). Figure~\ref{f:del-both} shows the results for \deltalogmbh\ as a function of redshift for both samples (RQQ only). Also shown in Figure~\ref{f:del-both} is \deltalogmbh\ averaged over redshift bins of $\Delta z = 0.1$ for each sample. Table~\ref{t:bins} shows various quantities averaged over these redshift bins. The mean \deltalogmbh\ is $+0.13$ for the HO3 sample, which indicates that the black hole mass implied by \sigthree\ and Equation \ref{e:tremaine} on average is less than the measured \mbh\ by 0.13 dex. Note that this heavily weights the abundant low-redshfit QSOs. We do not assign great significance to this overall offset, which may be within the uncertainties in the calibration of the formula for \mbh\ and of the use of \sigthree\ for \sigstar. The mean \deltalogmbh\ for the MO2 sample is $+0.41$ dex. 

Figure~\ref{f:del-both} shows an upward trend in \deltalogmbh, amounting to an increase of roughly 0.45~dex from low redshift to $z \approx 1$. Figure~\ref{f:mbh-both} shows the \mbhsigstar\ relationship with log \mbh\ and log $\sigma$ averaged over the same redshift bins as in Figure~\ref{f:del-both}. Figure~\ref{f:mbh-both} shows that the rise in \deltalogmbh\ with redshift results from an increase in \mbh\ not accompanied by a commensurate increase in \sigthree\ or \sigtwo. 

\subsection{Uncertainties}\label{s:uncert}

Does the trend Figure~\ref{f:del-both} represent a real dependence of the \mbhsigstar\ relationship on cosmic time? A number of uncertainties and selection effects bear discussion. 

\subsubsection{Slope of the Local \mbhsigstar\ Relationship}\label{s:slope}

Wyithe (2006) proposes a log-quadratic fit to the local \mbhsigstar\ relationship. For our range of \mbh, Wyithe's expression has a slope $d\, \mathrm{log}\,\mbh/d \mathrm{log}\,\sigstar = 4.5$, somewhat steeper than the slope of 4.0 in Equation \ref{e:tremaine}. The average \mbh\ in our redshift bins (Table \ref{t:bins}) increases by about 0.7 dex from the lowest three bins as a group to the highest three bins. Over the corresponding range in expected \sigstar, \mbh\ rises by $\sim0.07$ dex more in Wyithe's fit than in Equation \ref{e:tremaine}. If Wyithe's fit corresponds to the true \mbhsigstar\ relationship, our use of Equation \ref{e:tremaine} exaggerates the increase in \deltalogmbh\ with redshift by this amount. 

\subsubsection{Effect of the Radius--Luminosity Relationship}\label{s:rl}

The mean luminosity of our QSOs increases from log~\nuLnu\ = 44.0 for the lowest three redshift bins ($z \approx 0.25$) in Table \ref{t:bins} to 45.3 for the highest three ($z \approx 0.95$). A larger value of $\gamma$ in Equation \ref{e:mbh} would increase the \mbh\ derived for more luminous quasars, and amplify the trend of increasing \deltalogmbh\ with redshift shown in Figure~\ref{f:del-both}. The recent work of Bentz et al. (2006) argues against a $\gamma$ as large as 0.67 (Kaspi et al. 2005). If we used $\gamma = 0.60$, this would increase the trend in \deltalogmbh\ over the above redshift range by 0.12, that is, from +0.4 to +0.5 (prior to correcting for other biases).

\subsubsection{Accuracy Cuts}\label{s:rl}

Our restriction of our data set to the best quality measurements results in a severe winnowing of the final number of QSOs compared with the full number of QSOs in DR3. This includes a factor 2 (HO3) or 3 (MO2) reduction by visual inspections of the quality of the data and the fit. For the HO3 sample, there is already a trend of $\sim0.3$ in \deltalogmbh\ with redshift in the starting sample following the equivalent width error cut. The other numerical cuts have little effect on the trend, but the visual selection significantly raises the trend to the magnitude seen in Figure \ref{f:del-both}. For the MO2 sample, after the EW cuts, there is about 0.5 dex overall offset with little redshift trend. The FWHM cut brings this down to about 0.4 dex, and the other cuts have little effect. Finally, the visual selection lowers the offset at lower $z$, strengthening the trend in the figure. Thus, the overall sense of trend with redshift is present at all stages in the selection process, but it is substantially affected by the cuts made, including the visual inspections. Note however that the results prior to the inspections include a large percentage of objects with badly failed fits or other problems, which clearly should be excluded. 

\subsubsection{Broad Lines}\label{s:broad}

\begin{figure}[h]
\includegraphics[width=.45\textwidth]{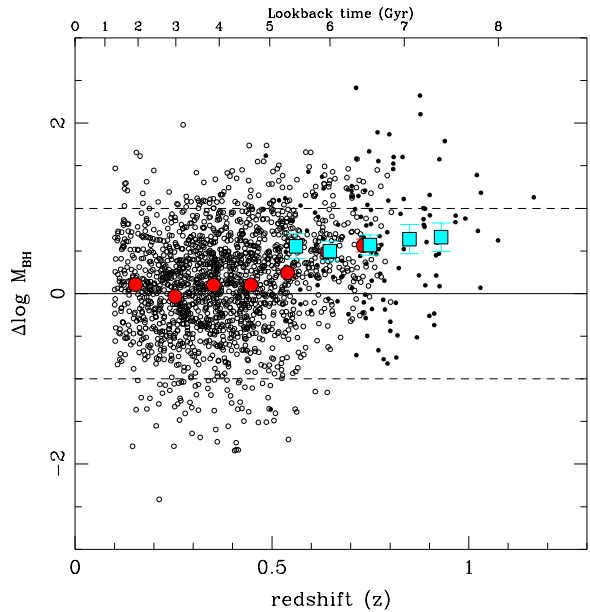}
\figcaption[ ]{Same as Figure~\ref{f:del-both}, but using recalibration to bring \mgii\ FWHM for the MO2 sample into agreement with \hbeta\ FWHM. Recalibration is based on a linear least squares fit to the data of Figure~\ref{f:hbmg}.
\label{f:del-both-cal}}
\end{figure}

We noted above that for larger \hbeta\ line widths, on average our measured \mgii\ widths are narrower than those of \hbeta\ (Figure \ref{f:hbmg}). If we use a linear fit to Figure~\ref{f:hbmg} to calibrate the widths of \mgii\ to bring them into agreement with those of \hbeta, this increases \mgii\ widths for FWHM $> 4000$~\kmps\ by $\sim0.1$ and \mbh\ by 0.2~dex. Using a linear least squares fit we find in units of \kmps: FWHM(\hbeta) $ = $ 0.645 FWHM(\mgii) $ + $ 890. Figure~\ref{f:del-both-cal} shows how the recalibration strengthens the rise of \deltalogmbh\ with redshift in Figure~\ref{f:del-both}, and gives a more consistent trend between HO3 and MO2. The recalibrated results give \deltalogmbh\ $ = 0.64$ and 0.66 for the bins at $z = 0.85$ and 0.93, respectively. The recalibration somewhat strengthens the trend of \deltalogmbh\ with $z$.

The HO3 and MO2 samples overlap for three redshift bins (Table~\ref{t:bins}). While there is general agreement between the two samples, the typical FWHM$_{\mathrm{BL}}$ (and thus \mbh) of the MO2 sample is systematically higher than that of the HO3 sample. This must result from object selection, because for individual objects Figure \ref{f:hbmg} shows, if anything, slightly smaller FWHM for \mgii\ than \hbeta. If usefully strong \oii\ correlates with wider \mgii, does this represent a potential bias? Boroson \& Green (1992) find significant correlations between optical properties of QSO, known collectively as ``eigenvector 1", including an anti-correlation between the strengths of \oiii\ and optical \feii\ emission. Hughes \& Boroson (2003) find a similar relationship for \oii\ strength. Likewise, in our sample objects with \oii\ strong enough to be detected have weaker \feii\ emission---optical as well as UV---than other objects. In eigenvector 1, \hbeta\ FWHM correlates with \oiii\ strength and anti-correlates with \feii\ strength. QSOs with detectable \oii\ will tend to have wider broad-line emission and, for a given luminosity, larger \mbh. This in turn is related to the Eddington ratio (Boroson 2003). The possibility of a resulting bias in \deltalogmbh\ should be kept in mind. Such a bias would require a systematic connection between a QSO's $L/\led$ and the ratio of \mbh\ to host galaxy mass, presumably involving the details of AGN fueling. (See \S 5.2 for discussion of biases involving the QSO luminosity.)

\subsubsection{Narrow Lines}\label{s:narrow}

We discussed above the removal of objects with \oiii$/$\hbetan\ $ \ll $ 10. If these objects systematically deviated from the \mbhsigstar\ relationship in a redshift-dependent way, their exclusion could bias the derived evolution. Note that this is not the same issue as the tendency for \oiii\ to weaken relative to {\it broad} \hbeta\ with increasing $L/\led$ (Boroson 2000). We examined 200 randomly selected objects that were rejected from our HO3 sample and found that 27 ($\sim 14$\%) of these were rejected for having \oiii$/$\hbetan\ $ \ll $ 10. We used the `splot' package in IRAF to directly measure the FWHMs of the broad-line component of \hbeta\ for these removed objects, and calculated their \mbh\ and \deltalogmbh. While the redshifts of these objects vary from $z \sim 0.1 - 0.5$ they mostly resemble their lowest redshift counterparts in terms of mean luminosity, \mbh, \deltalogmbh, and Eddington ratio). The mean Eddington ratio of the 27 objects is $\lbol/\led = -0.98$ and the mean \deltalogmbh\ $ = 0.04 \pm 0.11$. Reintegration of all such objects into the HO3 sample would have a negligible effect on our results.

We also noted above that for \oiii\ lines with peaky profiles our fits tended to crop the tops of the profiles. For roughly one in 10 QSOs, the \oiii\ lines were cropped by $> 10\%$ of the \oiii\ height, and were discarded from the sample. Curiously, this does not affect the magnitude or offset of the scatter in the \mbhsigstar relationship. The results for the heavily cropped objects showed that, if anything, they tended to cluster more closely around the local \mbhsigstar\ correlation than the non-cropped objects when using our automated (cropped) fits. Whatever causes the sharply peaked \oiii\ lines, our fits happen to crop in such a way as to recover an underlying profile similar in width to that of the non-cropped objects.

The tendency for spectra to become noisier with redshift means that, for a given equivalent width, wider \oiii\ and \oii\ lines are more difficult to measure. This increases the chance that higher-redshift QSOs with large \sigthree\ or \sigtwo\ will be rejected from the sample. To test the degree of bias introduced by this effect, we generated mock QSO spectra with noise to mimic our HO3 sample. For each object in the HO3 sample, we used the fitted line parameters (equivalent width and $\sigma_{\hbeta}$) to generate the \hbeta\ line with a pure Gaussian profile, and the measured 5100~\AA\ continuum flux to generate a power-law continuum of the form $\fnu \propto \nu^{-0.5}$ with a continuum noise level equal to the observed noise at 5100~\AA. The \oiii\ doublet was added using a pure Gaussian profile with the fitted \oiii\ equivalent width from the observed spectrum and a \sigthree\ drawn randomly from a Gaussian distribution with dispersion $\delta\sigthree = 0.13$~dex (Bonning et al. 2005) centered on a value inferred from the \mbh\ value for the object and Equation \ref{e:tremaine}. We then measured the mock spectra as though they were real data. This simulation was run several times, and we found a pattern of positive bias in \deltalogmbh\ that increases with redshift. For the HO3 redshift bins given in Table \ref{t:bins}, \deltalogmbh\ rises, on average, from 0.03~dex in the lowest redshift bin to 0.17~dex in the highest. The result of these simulations suggests that due to increasingly noisy spectra with redshift, our sample favors narrower \sigthree\ (and presumably \sigtwo) for a given \mbh. This bias can account for $\sim 0.1 - 0.2$~dex of the trend in \deltalogmbh\ in Figure~\ref{f:del-both}.

\subsubsection{Host Galaxy}\label{s:hostgal}

We have not removed the host galaxy contribution to the QSO spectra. We have estimated the galaxy contribution from the prominence of the Ca~II~K line in the spectra (Greene \& Ho 2006). We formed composite spectra, normalized to the continuum at 4000~\AA, of the objects in the various redshift bins of Table \ref{t:bins}. The Ca K line is conspicuous in the lowest redshift bin ($z = 0.15$), and becomes weaker with increasing redshift, as the QSOs become more luminous. We fit the $z = 0.15$ composite with a power-law continuum $F_\lambda \propto \lambda^{-1.56}$ combined with a typical elliptical galaxy spectrum. We used SDSS J141442.91$-$003236.8 at $z = 0.185$, which has $\sigstar \approx 220~\kmps$ based on our fits to the spectrum with a stellar template using a least squares fitting program. (We removed the narrow [Ne~III]~$\lambda3968$ and broad H$\epsilon$ lines from the composite on the basis of the $\lambda3869$ and H$\delta$ lines and standard intensity ratios. The subtraction of [Ne~III] and H$\epsilon$ has some effect on the red wing of the Ca K line but does not seriously affect the derived equivalent width.) A ratio of galaxy to power law of $\sim0.2$ reproduces the equivalent width of the Ca K line in the composite spectrum. By Equation \ref{e:mbh} this implies that we are overestimating the \mbh\ in this redshift bin by $\sim10\%$. This is insignificant for our conclusions. The EW of Ca~K suggests that the galaxy contribution becomes even less important with increasing redshift. This is consistent with the fact that $L/\led$ increases with redshift (Table \ref{t:bins}), implying an increase in $L/\lgal$ if $\lgal \propto \mbh$ (see Greene \& Ho 2006). Treu et al. (2004) found a $\sim50\%$ galaxy contribution in their SDSS AGN at $z = 0.37$. If the galaxy fraction in our lowest redshift bins were this large, we would be overestimating \mbh\ by $\sim0.15$~dex. The correction would be less at higher redhifts. Such a correction would increase by $\sim0.1$~dex the trend in \deltalogmbh\ with redshift found below. 

The effect of an underlying \hbeta\ absorption line in the galaxy component should be unimportant. Even for the E+A galaxies of Zabludoff et al. (1996), the Balmer line equivalent width is only a few \AA. The equivalent width of the broad \hbeta\ line is typically 100~\AA\ or more, and the galaxy contributes only a fraction of the total continuum.

\subsection{Luminosity Bias}\label{s:lumbias}

\begin{figure}[h]
\includegraphics[width=.45\textwidth]{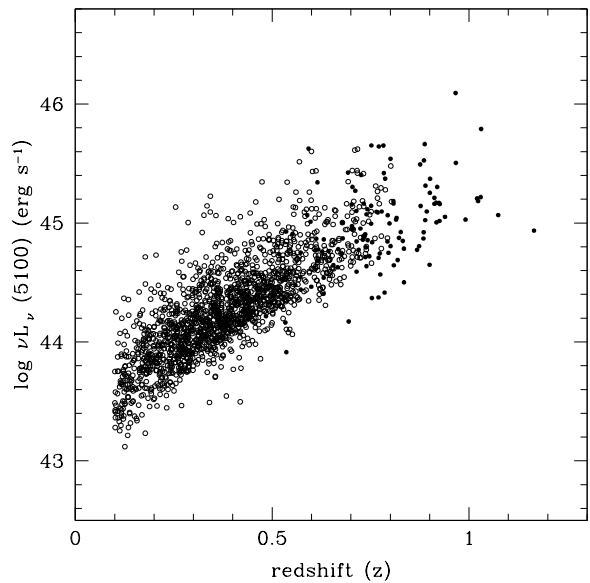}
\figcaption[lum-z-new.eps]{\nuLnu as a function of $z$. Open triangles are the HO3 data; closed squares are the MO2 data (excluding radio-loud QSOs).
\label{f:lum-z}}
\end{figure}

Another issue involves the effect of the limiting magnitude of the SDSS survey along with our quality cuts, which favor brighter objects. (Such a potential bias is mentioned by Treu et al. 2004.) This leads to a correlation between $L$, \mbh, and $z$ (see Table \ref{t:bins} and Figure~\ref{f:lum-z}). For nearby galaxies, Tremaine et al. (2002) find an rms dispersion of $\dmsigma = 0.3$ in log~\mbh\ at fixed \sigstar. If, for a given galaxy mass, galaxies with larger \mbh\ tend to have larger $L$, such galaxies will be over-represented in a flux-limited sample. This will lead to a positive Malmquist bias in the average \deltalogmbh\ for the sample. The bias will be stronger for a more steeply sloping galaxy mass function, which is starved for large galaxies. Because the mass function steepens for larger mass, the bias could increase with redshift. We give here a simple estimate of this bias as it affects our results. For the sake of concreteness, we illustrate the effect using dimensional quantities for \mbh, etc. However, our analysis could be expressed in an equivalent dimensionless form. We will argue that the magnitude of the bias depends on the typical luminosity of the sample QSOs at a given redshift, relative to the break in the slope of the QSO luminosity function at that redshift.

The galaxy mass function is often expressed in terms of a Schechter (1976) function
\begin{equation}
\label{e:schec}
\Phi(\mgal) = \Phi^* (\mgal/M_g^*)^{-a}\, e^{-\mgal/\mgalstar}.
\end{equation}
For redshifts $z \approx 1$, Drory et al. (2005) find $\mgalstar \approx 10^{11}~\msun$. For a typical ratio $\mbh/\mbulge = 0.0013$ (Kormendy \& Gebhardt 2001), this corresponds to $\mbh = 10^{8.1}~\msun$ if $\mbulge \approx \mgal$. Table \ref{t:bins} shows that the average \mbh\ in our redshift bins progresses from below to above $10^{8.1}~\msun$ with increasing redshift. Therefore, the possibility that the trend in Figure~\ref{f:del-both} involves such a bias is a concern.

For a rough estimate of this bias, we carried out simple Monte Carlo trials. Values of $\mgal$ were distributed as in Equation~\ref{e:schec} with $\mgalstar = 10^{11}~\msun$, covering a wide mass range around \mgalstar. For each galaxy, \mbh\ was drawn randomly from a Gaussian distribution in log~\mbh\ centered on $\mbh = 0.0013~\mgal$ with an rms dispersion in log~\mbh\ of \dmsigma. For this \mbh, the luminosity was drawn from a uniform random distribution in $L$ (not log$L$) between $L/\led = 0.001$ and 0.3. This crudely simulates our typical values of $L/\led$; see below for alternative assumptions. The resulting QSO luminosity function mirrors the galaxy mass function, with a break at $\lbreak \approx 0.5 L_0$, where $L_0 = 0.3\led(\mbhstar)$ and $\mbhstar = 0.0013 \mgalstar$. Roughly speaking, we expect that at a given redshift, QSOs brighter than some minimum luminosity will have sufficient S/N to be retained in our samples. For higher redshifts, this minimum luminosity will increase. We simulate this situation by selecting from our Monte Carlo runs those QSOs with $L > \lcut$, where \lcut\ is an adjustable parameter. The effect of increasing \lcut\ should indicate how the bias varies with increasing redshift, because the higher redshift objects are typically more luminous. In a given simulation, for each of several values of \lcut, we computed average values of log~\mgal, log~\mbh, log~$L$, and \deltalogmbh\ for the kept objects, i.e., those having $L > \lcut$. For the models, we have $\deltalogmbh = $ log \mbh\ - log 0.0013 \mgal. Table \ref{t:bias} shows the results of these simulations for $\dmsigma = 0.3$ and 0.5. As expected there is a bias in the sense of positive \deltalogmbh\ that increases with $\lcut/L_0$, reaching values as high as $\deltalogmbh \approx 0.5$ for $\dmsigma = 0.3$ and 0.8 for $\dmsigma = 0.5$. This bias has its origin in the dispersion of the black hole mass -- bulge relationship and is roughly proportional to \dmsigma.

Motivated by the observed QSO luminosity function (Boyle et al. 2003) discussed below, we also computed models with an alternative form of the galaxy mass function,
\begin{equation}
\label{e:marshall}
\Phi(L) = \Phi^* [(\mgal/\mgalstar)^{-a} + (\mgal/\mgalstar)^{-b}]^{-1}
\end{equation}
with $a = -3.41$, $b = -1.58$. The bias is similar for the two different functional forms of the mass function, when expressed as a function of $\lcut/\lbreak$ (see Table \ref{t:bias}).

How can we relate these simulations to our observed QSOs? Consider a model in which the probability of fueling the black hole, $p(\mbh)$, decreases with increasing \mgal; but when fueling does occur, it follows our assumed probability distribution for $L/L_{\mathrm{Ed}}$. This will have the same effect on the observed number of QSOs at various values of $L$ and \mbh\ as if we had assumed a modified galaxy mass function $\Phi^\prime(\mgal) = p(\mbh)\,\Phi(\mgal)$, where $\mbh = 0.0013\mgal$. This suggests that what matters for the bias is the slope of the QSO luminosity function. As the slope becomes steeper with increasing $L$, the bias will increase. A practical approach is therefore to compare our typical luminosity with the break in the QSO luminosity function, redshift by redshift.

Boyle et. al (2003) give the $B$-magnitude luminosity function as 
\begin{eqnarray}
\label{e:boyle}
\Phi(M_B) = \Phi^*_M \{ \mathrm{dex}[0.4\Delta M_B(\alpha+1)]\\
\nonumber + \mathrm{dex}[0.4\Delta M_B(\beta+1)] \}^{-1}
\end{eqnarray}
with $\alpha = -3.41$, $\beta = -1.58$, $\Delta M_B \equiv M_B - M^*_B(z)$, and $M^*_B(z) = -21.92 - 2.5(k_1 z + k_2 z^2)$ adjusted for our \hnot. Equation~\ref{e:boyle} is the magnitude version of Equation~\ref{e:marshall} for luminosity rather than mass, and it is the motivation for our use of Equation~\ref{e:marshall}. Our $z = 0.9$ to 1.0 bin has a mean log~\nuLnu of 45.25. For this redshift, the break magnitude in the Boyle et al. luminosity function is $M^*_B = -24.50$, corresponding to $\mathrm{log}~\nuLnu = 45.28$ for a typical QSO continuum slope of $L_\nu \propto \nu^{-0.44}$ (Vanden Berk et al. 2001). Thus, our average luminosity of the QSOs in this redshift bin is 0.03 dex fainter than $L^*$ for the observed luminosity function at this redshift. We therefore consider simulations using Equation~\ref{e:marshall} with a value of \lcut\ adjusted so that the mean luminosity of the kept QSOs ($L > \lcut$) is 0.03 dex fainter than $L_*$ for the model luminosity function. This sample of kept objects has an average $\deltalogmbh \approx 0.21$ for $\dmsigma = 0.3$. For the lowest redshift bin in Table 1 ($z = 0.1$ to 0.2) we similarly find a bias of $\sim 0.11$. Therefore, this bias can account for a {\em rise} in \deltalogmbh\ of $\sim0.1$ from $z = 0.1$ to $z = 1$. This increases to a differential bias of $\sim0.2$ if we double the value of \dmsigma.

We also ran simulations in which all model QSOs had $L/\led = 0.3$ rather than ranging down to 0.001 as above. (In this case $\lbreak \approx L_0 = 0.3 \led(\mbhstar)$). The bias in \deltalogmbh\ was only $\sim 0.02$ larger in this case, when expressed in terms of $\lcut/L_0$. These and other trials show that the details of the $L/\led$ distribution does not seriously affect the bias, as long as the selected model sample bears a given relationship to the break in the model luminosity function. It is in this sense that the simulation is essentially scale free, as noted above.

The bias considered here is, however, sensitive to the role of galaxy mass in AGN fueling. The discussion above assumes a distribution in $L/\led$ that is the same at each value of \mgal. Thus, fueling is proportional to \mbh, with some probability function for $\dot{M}/\mbh$. This means that a galaxy with a large black hole for its mass is more likely to be seen as a QSO in our sample. Suppose, however, that QSO fueling is instead governed by galaxy mass, such that $L$ is given by some probability function for $\dot{M}/\mgal$, independent of \mbh. Now the QSO luminosity does not favor objects with large \mbh\ relative to \mgal, even though there is still a distribution in \mbh\ for a given \mgal. Consequently, we expect no bias. We conducted simulations in which $L/\mgal$ followed a linear probability distribution from a maximum value $L/\led(0.0013\mgal) = 0.3$ or 1.0 down to small values. These trials gave zero bias, within the Monte Carlo noise. An intermediate case would involve fueling tied to \mgal\ but a luminosity bounded by the Eddington limit. Table \ref{t:bias} gives results for two such simulations. For $ L = 0.3\led(0.0013\mgal)$, few black holes are so small that the Eddington limit comes into play, and the bias is essentially zero. For $L = 1.0\led(0.0013\mgal)$, the Eddington limit affects many of the QSOs (those with smaller than average \mbh), and the bias is significant but not so large as in models with fueling proportional to \mbh\ and the same average $L/\lbreak $.

The simulations show some bias even in the lowest redshift bin, which may bear consideration in studies attempting to calibrate the prescription for \mbh\ in Equation~\ref{e:mbh} by using the \mbhsigstar\ relationship (Onken et al. 2004). Our lowest redshift bins for the observed QSOs in fact have $\deltalogmbh \approx 0.1$, consistent with our bias estimate.

We conclude that, due to the scatter in the \mbhsigstar\ relationship, observational selection favors the brighter QSOs, and hence the bigger black holes, for a given galaxy mass. This bias can account for $\sim0.1$ of the trend in \deltalogmbh\ in Figure~\ref{f:del-both}. The scatter in the \mbhsigstar\ relationship \dmsigma\ is measured fairly well in the mass range of our observed QSOs, and seems unlikely to be as much as twice the adopted value of 0.3. The bias appears not to be sensitive to the exact form the the QSO luminosity function, judging from the two cases considered above, nor is it sensitive to the detailed $L/\led$ distribution. However, the bias is sensitive to whether QSO fueling is governed by \mbh\ or \mgal. In the latter case, the bias is reduced, and we derive a larger evolution in the \mbhsigstar\ relationship.

\subsection{Conclusion}

We have identified two significant biases involving the spectrum noise and luminosity selection effects. Since these biases arise from two unrelated phenomena---the scatter in the \sigthree--\sigstar\ relationship and the scatter in the \mbhsigstar\ relationship ---we may add them linearly to estimate their cumulative effect on the mean \deltalogmbh. We find that much of the overall trend in Figure~\ref{f:del-both} can be attributed to these selection effects. Specifically, there is a rise of $\sim0.4$ in \deltalogmbh\ from our lowest three redshift bins to the highest three. Our nominal results for the spectrum noise bias of $\sim0.15$ and the luminosity-driven bias of 0.1 combine to give a total bias of 0.25, leaving a derived evolution of $\sim0.2$ in \deltalogmbh. The noise bias presumably is well determined. An increase in \dmsigma\ with redshift could contribute to the measured evolution, but this would still be a form of evolution of the \mbhsigstar\ relationship. If QSO fueling is driven by galaxy mass, the luminosity bias is reduced, and the derived evolution becomes $\sim0.3$. Alternatively, if we recalibrate the \mgii\ width as discussed above, the derived evolution again becomes $\sim0.3$.

\section{Discussion}

After accounting for selection effects, we find some evolution in the \deltalogmbh\ relationship in the redshift range $z = 0.1 - 1.2$, such that black holes are too large for a given bulge velocity dispersion. In contrast, S03 found no systematic evolution in \deltalogmbh\ out to redshift $z \approx 2 - 3.$ The reason for this difference is unclear. S03's high redshift objects had log \mbh\ $\sim 9.5$ while our $z \approx 1$ objects typically have log~$\mbh \approx 8.5$. SO3's high redshift QSOs have very wide \oiii\ lines, with $\sigthree \approx 500~\kmps$. However, the sample was small and the spectral resolution marginal.

Treu et al. (2004) report measurements of \sigstar\ in a small sample of SDSS AGN at redshift $z \sim 0.37$. They find $\Delta$~log~$\sigma = -0.16$, corresponding to $\deltalogmbh = +0.64$. (Excluding one outlier with large measurement errors in \sigstar, their result becomes $\Delta$~log~$\sigma = -0.12$.) There is no such offset in the low redshift AGN sample that they use for comparison (Merritt \& Ferrarese 2001). While noting the need for a larger sample, Treu et al. raise the possibility of evolution in the \mbhsigstar\ relationship. Our results show no offset of this magnitude in the redshift range of Treu et al.'s study. Using the SDSS spectra, we have measured the \oiii\ line widths of the six SDSS objects for which Treu et al. quote a value of \sigstar. On average, \sigthree\ is larger than their quoted \sigstar\ by 0.09 dex, and in particular the outlier in \sigstar\ has a \sigthree\ in agreement with the local \mbhsigstar\ relationship. Perhaps this is an indication that Treu et al.'s measurements of \sigstar\ are affected by contributions to the starlight from relatively face-on galactic disks, a possibility that Treu et al. note. We also find that a direct measurement of the FWHM of the broad \hbeta\ line in Treu et al.'s SDSS objects is typically narrower by about 0.1 dex than the FWHM implied by Treu et al.'s measurements of the second moment of the \hbeta\ profile (i.e., the ``rms'' multiplied by 2.35). As a result, the offset in \deltalogmbh\ resulting from our measurements of the Treu et al. objects is considerably less than their result. It will be important to obtain high S/N measurements of \sigstar\ for a larger sample of AGN.

Using infrared spectra by Sulentic \etal\ (2004), Shields et al. (2006b) analyze \mbh\ and \sigthree\ for nine QSOs at an average redshift of $z = 1.3$ and find a mean $\deltalogmbh = 0.3$. This resembles our result, but involves somewhat larger black holes with a mean log $\mbh = 9.3$. Peng et al. (2006) summarize measurements of QSO host galaxy magnitudes for lensed QSOs at redshifts up to $z \approx 4.5$. After allowance for evolution of the host galaxy stellar population, Peng et al. find an upper limit of a factor of 2 evolution in the \mbh/\mgal\ mass ratio for $1 < z < 1.7$, consistent with our result. At $z > 1.7$, the black holes are too large for the host galaxies by a factor of 4. This assumes an evolution of the \mbh-\lbulge\ relationship similar to that of the \mbhsigstar\ relationship. Both results suggest that BHs were larger relative to the bulge in the past. The host galaxies must have acquired much of their present-day mass after growth of the black hole was largely completed. Croton (2006) presents simulations of galaxy and black hole growth in which he finds a decrease in $\mbh/\mbulge$ with time. His ``dynamic'' model shows a decrease in $\mbh/\mbulge$ by a factor of two from $z = 1$ to the present, similar to our observational result. This involves growth of the bulge by late-time mergers that do not significantly fuel black hole growth. 

This scenario differs from the evolutionary sequence modeled by Di Matteo et al. (2005; see also Robertson \etal\ 2006), who compute mergers of disk galaxies in which most of the stars are already formed. The modest initial black holes grow by accretion of gas until the AGN luminosity expels the residual gas. The expulsion terminates black hole growth and gives a remarkably tight \mbhsigstar\ relationship, but in this case the black hole catches up to the stellar mass of the galaxy. The need for the QSO luminosity to deposit sufficient energy in the gas to liberate it from the host galaxy gravitational potential gives a plausible connection between \mbh\ and \sigstar\ that does not play the same role if the host galaxy must catch up to that of the black hole. 

Shields et al. (2006a) have investigated the \mbhsigstar\ relationship in high redshift QSOs using the width of the radio CO lines as a probe of the galactic potential. Taking $\sigma_\mathrm{CO} \equiv \mathrm{FWHM(CO)}/2.35$ and \mbh\ derived from the width of the broad Mg~II and C~IV lines, they find $\deltalogmbh = 1 - 2$ for $z = 4 - 6$. These are large black holes with $\mbh \approx 10^{9.5} $ to $10^{10}$ solar masses and comparatively modest host galaxies. (An uncertainty is the possibility of relatively face-on CO disks in these QSOs, as noted by Shields et al. 2006a and Carilli \& Wang 2006.) This is in the same qualitative sense as offset from the local \deltalogmbh\ relationship found by Peng et al. (2006) for their $z > 1.7$ QSOs. However, Shields et al. (2006a) argue that, in general, the host galaxies of their high redshift quasars will never catch up to the expected mass for their \mbh, because the present day galaxy luminosity function does not contain sufficient numbers of galaxies of the necessary size (Shields \etal\ 2006b). While the hosts of the $z > 4$ QSOs of Shields et al. (2006a) may experience some later growth, these extreme black holes evidently represent a breakdown in the \mbhsigstar\ relationship observed locally for smaller masses.

\section{Summary}

We have used the SDSS DR3 to assess how well QSOs up to redshift $z \approx 1$ follow the \mbhsigstar relationship for nearby galaxies. We created two data samples: one consisting of objects with the optical emission lines \hbeta\ and \oiii\ in the approximate redshift range $0.1 < z < 0.8$ to compare with results of previous studies (the HO3 sample); the other sample consisting of objects with rest-frame ultraviolet lines, \mgii\ and \oii, with redshifts $z < 1.2$ (the MO2 sample). 

A summary of our findings is as follows:

\begin{enumerate}

\item
The widths of \oiii\ and \oii\ show overall agreement, with a mean log \sigthree\ - log \sigtwo\ of -0.013 and a dispersion of 0.12.

\item
There is generally good agreement between the widths of \hbeta\ and \mgii, though for wider \hbeta\ lines, \mgii\ tends to be narrower than \hbeta. The mean log FWHM(\hbeta) - log FWHM (\mgii) is 0.05.

\item
\mgii\ and \oii\ can be used to study the \mbhsigstar\ relationship up to redshifts $z \approx 1.2$. There is little evolution in the \mbhsigstar\ relationship between now and redshift $z \sim 0.5$. For redshifts $z > 0.5$ there is a trend in \deltalogmbh\ with redshift in the sense that bulges are too small for their black holes. Part of this trend results from selection effects. Accounting for bias, we find 0.2 dex evolution in the \deltalogmbh\ relationship between now and redshift $z = 1$, corresponding to a time when the universe was approximately six billion years old. Evolution of this nature is predicted by some models of galaxy and black hole growth.

\end{enumerate}

\acknowledgments

We thank Sandy Faber, Gary Hill, Eliot Quataert, and Bev Wills for helpful discussions, and Ashley Davis, Pamela Jean, and Michael Shields for assistance. G.A.S. gratefully acknowledges the hospitality of Lick Observatory and the support of the Jane and Roland Blumberg Centennial Professorship in Astronomy. This material is based in part on work supported by the Texas Advanced Research Program under grant 003658-0177-2001; by the National Science Foundation under grant AST-0098594; and by NASA under grant GO-09498.04-A from the Space Telescope Science Institute, which is operated by the Association of Universities for Research in Astronomy, Inc., under NASA contract NAS5-26555. E.W.B. is supported by Marie Curie Incoming European Fellowship contract MIF1-CT-2005-008762 within the 6th European Community Framework Programme. We thank the SDSS team for the enormous effort involved in conducting the survey and making the results conveniently accessible to the public.

Funding for the Sloan Digital Sky Survey (SDSS) has been provided by the Alfred P. Sloan Foundation, the Participating Institutions, the National Aeronautics and Space Administration, the National Science Foundation, the U.S. Department of Energy, the Japanese Monbukagakusho, and the Max Planck Society. The SDSS Web site is http://www.sdss.org/. The SDSS is managed by the Astrophysical Research Consortium (ARC) for the Participating Institutions. The Participating Institutions are The University of Chicago, Fermilab, the Institute for Advanced Study, the Japan Participation Group, The Johns Hopkins University, the Korean Scientist Group, Los Alamos National Laboratory, the Max-Planck-Institute for Astronomy (MPIA), the Max-Planck-Institute for Astrophysics (MPA), New Mexico State University, University of Pittsburgh, University of Portsmouth, Princeton University, the United States Naval Observatory, and the University of Washington.

\clearpage

\begin{deluxetable}{lcccccc}
\tablewidth{0pt}
\tablecaption{Average Quantities for Redshift Bins\label{t:bins}}
\tablehead{
\colhead{$z$ (\# of objects)}          &
\colhead{\nuLnu}       &
\colhead{$\sigma_{\mathrm{NL}}$} &
\colhead{\mbh}         &
\colhead{\deltalogmbh} &
\colhead{$\lbol/\led$} &
\colhead{FWHM$_{\mathrm{BL}}$} \\
\colhead{ }            &
\colhead{(\ergps)}     &
\colhead{(\kmps)}      &
\colhead{(\msun)}      &
\colhead{ }	       &
\colhead{ }            &
\colhead{\kmps}}
\startdata
{\it HO3 sample} & & & & & & \\
0.15 (213) & 43.77 & 2.14 & 7.59 & 0.11 & -0.96 & 3.52 \\
0.25 (344) & 44.01 & 2.21 & 7.73 & -0.04 & -0.87 & 3.53 \\
0.35 (403) & 44.23 & 2.22 & 7.90 & 0.10 & -0.82 & 3.56 \\
0.45 (332) & 44.40 & 2.24 & 7.98 & 0.10 & -0.73 & 3.55 \\
0.54 (200) & 44.60 & 2.23 & 8.10 & 0.24 & -0.65 & 3.56 \\
0.64 (105) & 44.75 & 2.21 & 8.25 & 0.49 & -0.65 & 3.59 \\
0.73 (47) & 45.02 & 2.22 & 8.37 & 0.56 & -0.50 & 3.59 \\
\\
{\it MO2 sample} & & & & & & \\
0.56 (14) & 44.68 & 2.29 & 8.45 & 0.36 & -0.91 & 3.71 \\
0.65 (18) & 44.78 & 2.25 & 8.30 & 0.37 & -0.67 & 3.61 \\
0.75 (43) & 44.97 & 2.27 & 8.45 & 0.43 & -0.62 & 3.64 \\
0.85 (25) & 45.01 & 2.26 & 8.48 & 0.50 & -0.61 & 3.66 \\
0.93 (14) & 45.25 & 2.27 & 8.55 & 0.53 & -0.45 & 3.62 \\
\enddata
\tablecomments{Excludes radio-loud QSOs. All quantities are in log units except for redshift. $\sigma_{\mathrm{NL}}$ denotes \sigthree\ for the HO3 sample and \sigtwo\ for the MO2 sample; FWHM$_{\mathrm{BL}}$ denotes the \hbeta\ FWHM for the HO3 sample and \mgii\ FWHM for the MO2 sample. Bins with fewer than ten objects were excluded.}

\end{deluxetable}

% Begin Table 2

\begin{deluxetable}{lccccc}
\tablewidth{0pt}
\tablecaption{Results of Bias Simulations\label{t:bias}}
\tablehead{
\colhead{$\mathrm{log}$}   &
\colhead{$\mathrm{log}$}   &
\colhead{$\mathrm{log}$}   &
\colhead{$\mathrm{log}$}   &
\colhead{$\Delta\,\mathrm{log}$}       &
\colhead{$\mathrm{log}$}  
\\
%second row of column heading
\colhead{$\lcut/L_0$}     &
\colhead{\mgal}         &
\colhead{\mbh}          &
\colhead{\nuLnu}       &
\colhead{\mbh }	         &
\colhead{$L$/\lbreak } 
}
\startdata
{\it Schechter, $\dmsigma = 0.3$} & & &  & & \\
  -1.37  & 10.35  &  7.57 &  -2.07  &  0.10  &  -0.51 \\
  -0.67   & 10.71  &  8.02  & -1.57  &  0.19  &  -0.01 \\
   0.00  & 11.05  &  8.50  & -1.03  &  0.34   &  0.53 \\
   0.48  & 11.33  &  8.93 &  -0.61  &  0.49  &   0.95 \\
\\
{\it Schechter, $\dmsigma = 0.5$} & & &  & & \\
  -1.37  & 10.32  &  7.68 &  -1.98  &  0.25  &  -0.42 \\
  -0.67  & 10.58  &  8.13  & -1.47   & 0.44  &   0.09 \\
   0.00  & 10.84  &  8.62  & -0.95  &  0.67  &   0.61 \\
   0.48  & 11.05   & 8.99  & -0.57  &  0.83   &  0.99 \\
\\
{\it Boyle, $\dmsigma = 0.3$} & & & & & \\
  -1.37 &  10.27  &  7.50 &  -2.13   & 0.12  &  -0.57 \\
  -0.67  & 10.65  &  7.98  & -1.60  &  0.21  &  -0.04 \\
   0.00  & 11.01  &  8.50  & -1.04  &  0.37   &  0.52 \\
   0.48  & 11.33  &  8.92 &  -0.59  &  0.48  &   0.97 \\
 \\
{\it Boyle, $\dmsigma = 0.3$}, $L = 0.3\led(0.0013\mgal)$ & & & & & \\
  -0.40  &  10.85  &  7.96  &  -1.41  &  0.00  &  -0.15\\
   0.48  &  11.63   &  8.76  &  -0.63  &  0.00  &   0.63\\
  \\ 
  {\it Boyle, $\dmsigma = 0.3$}, $L = 1.0\led(0.0013\mgal)$ & & & & & \\  
   -0.40   &  11.00  &  8.21  &  -0.92  &  0.09  &   -0.18\\

\enddata

\tablecomments{
Average quantities for model QSOs satisfying $L>\lcut$ in simulations (see text). Simulations are defined by the assumed form of the galaxy mass function (Schechter or Boyle) with characteristic mass \mgalstar, the dispersion \dmsigma\ in the \mbhsigstar\ relationship, and the value of $\lcut/L_0$, where $L_0 = 0.3\led(\mgalstar)$. Simulations are related to observed QSOs through the quantity $L/\lbreak$, where $\lbreak \approx 0.5 L_0$ is the luminosity at the break in the model QSO luminosity function, as explained in the text. Lower entries assume QSO luminosity proportional to \mgal, but limited to \led(\mbh).
 }

\end{deluxetable}

\end{document}